\documentstyle[eqsecnum,pre,preprint,aps,epsfig]{revtex}

\tightenlines

\newcommand{\bq}{\begin{equation}}
\newcommand{\eq}{\end{equation}}
\newcommand{\bqa}{\begin{eqnarray}}
\newcommand{\eqa}{\end{eqnarray}}

\def\prep#1#2#3{ Phys. Rep. ${\bf{#1}}$ (#2) #3}
\def\etal{{\it et.al.\/}}

\begin{document}

\draft
\preprint{PM/01-71}

\title{SUSY Scalar Production in the Electroweak Sudakov Regime
of Lepton Colliders
\footnote{Partially supported by EU contract HPRN-CT-2000-00149}}

\author{M. Beccaria$^{a,b}$, M. Melles$^{c}$,
F.M. Renard$^d$ and C. Verzegnassi$^{e, f}$ \\
\vspace{0.4cm}
}

\address{
$^a$Dipartimento di Fisica, Universit\`a di
Lecce \\
Via Arnesano, 73100 Lecce, Italy.\\
\vspace{0.2cm}
$^b$INFN, Sezione di Lecce\\
\vspace{0.2cm}
$^c$ Paul Scherrer Institute (PSI), CH-5232 Villigen, Switzwerland \\
\vspace{0.2cm}
$^d$ Physique
Math\'{e}matique et Th\'{e}orique, UMR 5825\\
Universit\'{e} Montpellier
II,  F-34095 Montpellier Cedex 5.\hspace{2.2cm}\\
\vspace{0.2cm}
$^e$
Dipartimento di Fisica Teorica, Universit\`a di Trieste, \\
Strada Costiera
 14, Miramare (Trieste) \\
\vspace{0.2cm}
$^f$ INFN, Sezione di Trieste\\
}

\maketitle

\begin{abstract}

We consider the production of SUSY scalar pairs at lepton colliders,
for c.m. energies much larger than the mass of the heaviest SUSY (real
or virtual) particle involved in the process. In that energy regime, we
derive the leading and subleading terms of the electroweak Sudakov
logarithms in the MSSM, first working at one loop with physical states
and then resumming to all orders with asymptotic expansions. We show
that the first order of the resummed expression reproduces the physical
one loop approximation, and compute systematically the possible effects
on various observables both at one loop and to all orders. We discuss
the regimes and the observables where the one loop approximation can or
cannot be trusted, working in an energy range between 1 TeV and 4 TeV under a
"light" SUSY mass assumption. As a byproduct of our analysis, we
propose a determination of the MSSM parameter $\tan\beta$ 
showing how a relative accuracy $\simeq 25\%$ can be easily achieved 
in the region $\tan\beta \gtrsim 14$, 
under reasonable experimental assumptions.

\end{abstract}


\section{Introduction}

The fact that the process of electron-positron annihilation into a
Standard Model (SM) fermion pair is dominated at high energy by large
electroweak logarithms of ``Sudakov-type" ~\cite{1} has become nowadays
completely established. After the first numerical calculations at one
loop, that evidentiated the unexpectedly large size of the leading
quadratic (DL) and subleading linear~\cite{2,3} (SL) terms, a
great amount of theoretical work has been devoted to the task of
computing this type of effect beyond the one-loop order. This hard
calculation becomes imperative if one wants to provide a safe
theoretical description of the process for c.m. energies in the TeV
range, where the truncation at one loop of the perturbative expansion
would be certainly unreliable~\cite{2}.

In general, one can say that two types of approaches have been
followed. The first one is based on the study of the process in a
suitable asymptotic energy region, where either technical or
theoretical simplifications are supposed to be valid. Although this
definition is not compulsory, we shall refer to this energy range as
that where an ``electroweak Sudakov regime'' has settled. In this
range, a resummation of the Sudakov logarithms to all orders
has been proposed by different groups~\cite{4}. For the specific case
of a final fermion pair production, the results seem nowadays
to agree, as
thoroughly discussed in a very recent paper~\cite{5}, and for a
detailed comparison we defer to the existing literature~\cite{4,6}.

In the second approach, one has tried to make explicit calculations of
the Sudakov effects at two loops, working with ``physical'' (i.e. not
asymptotic) processes and computing the high energy limits of the
derived expressions~\cite{7}. This second approach clearly provides a
very important test of the reliability of the first one, by comparison
of the corresponding two-loop contributions. At the moment, this
comparison has been successful for the leading logarithmic terms of
the asymptotic expansion. For the subleading terms, 
which depend both on the
c.m. energy $\sqrt{q^2}$ and the c.m. scattering angle $\theta$, a
clean high energy resummation prescription now
exists~\cite{5}, but the corresponding ``physical'' two-loop
calculation is not yet available. In spite of this, a fair
conclusion is that, at the moment, the electroweak Sudakov logarithms
in the process of electron - positron annihilation are under
control, provided that the simple one-loop expansion is replaced, in
the TeV c.m. energy range, by a more complete calculation. To be more
precise, it should be stressed, though, that the necessity of this
replacement is strongly dependent on the considered observable and
energy, and that for a special set of quantities (like, e.g.,
forward-backward asymmetries) the validity of a one loop expansion
might be still acceptable at realistic collider energies. For an
exhaustive discussion of this point we defer to a very recent
paper~\cite{8}.

An important question in this subject is that of when the
``electroweak Sudakov regime'' starts. Otherwise stated, at which
energy can one assume that an expansion of Sudakov type provides a
``satisfactory'' description of the various processes, i.e. one that
reproduces the relevant experimental observables at the requested
theoretical accuracy? In~\cite{3}, a tentative one loop analysis was
prepared by fitting the numerical values at variable energy of various
quantities, rigorously computed, with a logarithmic expansion ``\`a la
Sudakov'' that included, beyond the leading and the sub-leading terms
(including the RG ones) an unknown constant, to be fitted. The result
showed that, in the SM, an ``electroweak Sudakov regime'' was actually
settling when the c.m. energy approached the typical value $\sqrt{q^2}
= 1 \mbox{TeV}$, since the fitted parameters of the logarithms where
exactly the theoretical Sudakov ones. It also showed that, at such
energies, the numerical value of the constant term, although smaller
than those of the various logarithms, was not negligible at the
typical level of accuracy of a relative one percent (to be assumed
from now on as the aimed experimental reach of future lepton
colliders). The conclusion thus seems to be that, within the SM,
electron - positron annihilation appears to enter a ``Sudakov regime''
for c.m. energies of the TeV size to be hopefully reached by a future
LC~\cite{LC} or CLIC~\cite{10} experiments.

As a remark that appears, at this stage, somewhat formal, it can be
finally noticed that the knowledge of the Sudakov logarithms (and,
also, of the well known and precisely determined RG ones~\cite{RG})
would be sufficient to predict the \underline{slope} (i.e. the
variation with energy) of each
experimental observables in its ``Sudakov regime'' since clearly, in
this calculation, the constant terms will disappear. Within the SM,
this procedure would not provide particularly interesting
consequences, although it could be always considered a valuable
overall test of the hard theoretical calculations.

All the previous statements and conclusions are strictly valid, as we
said, in the SM theoretical framework. A spontaneous question that
arises at this point is that of whether the obtained picture remains
valid in a theoretical extension of the SM, for which the same type of
perturbative expansion is supposed to be valid. A rather natural first
candidate of this kind seems to be the minimal supersymmetric extension
of the SM, the MSSM. Analogously, the rather natural processes to be
considered seem to be the same ones that we have previously listed,
for which the Sudakov effect in the SM has been fully computed. Here,
the extra particles of the MSSM might induce new virtual effects at
the one-loop level, thus modifying the coefficients of both
the leading and the sub-leading Sudakov logarithms (and, of course, and
in an already known way, also those of the RG ones).

An important preliminary problem that arises at this point is that of
determining the c.m. energy at which the ``SUSY electroweak Sudakov
regime'' starts. The latter energy is fixed by the request of being
sufficiently larger than the mass of the heaviest SUSY particle that
appears in the diagrams that generate Sudakov logarithms. Clearly, no
precise answer can be given at the moment to this question. In spite
of this shortage, one can still proceed in a correct general way by
first defining as $M$ the unknown heaviest relevant SUSY mass and
computing the Sudakov expansion at asymptotic
energies $\sqrt{q^2} \gg M$.
A next, more speculative step, would be that of assuming a ``light
$M$'' situation characterized by ``reasonable'' $M$ values, say below
the TeV range. This would allow to make detailed numerical predictions
in a supposedly valid SUSY Sudakov regime, that might be easily
modified as soon as supersymmetric particles were finally detected.

The first investigation of SUSY Sudakov effects in the process of
electron - positron annihilation into a (charged) fermion -
anti-fermion pair was recently performed at the perturbative one-loop
level~\cite{13}, and we defer the interested reader to that reference
for a detailed discussion of the various considered processes and
observables. The two main general results that were derived are the
fact that virtual SUSY exchanges in those processes \underline{do}
generate Sudakov logarithms and the fact that these logarithms are of
sub-leading (SL) kind and in the generally adopted definition,
``universal''. At the one loop level, working in the 't Hooft $\xi=1$
gauge, they are only generated by vertex diagrams and \underline{not}
by diagrams of box type. Therefore, they are independent of $\theta$,
the c.m. scattering angle. From a formal point of view, they are
strictly analogous to the sub-leading logarithms of Yukawa type that
arise in the SM for final massive quark production, in practice for
production of third family quark-antiquark pairs.

For a more detailed numerical description, it was assumed in~\cite{13}
that the value of the heaviest SUSY mass $M$ was equal to
(approximately) a few hundred GeV. Under this assumption, the
numerical effects of the SUSY SL at one loop began to be appreciable
(say, of a relative few percent) when the c.m. energy was in the TeV
range, in particular they were definitely visible in the supposed CLIC
($\sqrt{q^2}\simeq 3$ TeV) region, where in conclusion the virtual
effects of the MSSM, under this ``light $M$'' working assumption,
would be clearly experimentally testable.

An important feature to be clarified at this point is that of which
information on certain SUSY parameters can be achieved in this way. In
general, the situation is complicated for two reasons: in first
place, one does not know exactly the value of $M$; secondly, the role
of next-to-sub-leading constant SUSY terms in an asymptotic
expansion is unknown (note that, differently from the SM case, a
numerical fit of the constant term like that performed in~\cite{3}
would require the knowledge of all the SUSY parameters that might
enter this quantity, and results thus, in practice, hardly
performable).

A possible approach that gets rid of the two previous difficulties has
been very recently proposed~\cite{14}. It is based on the observation
that, in the calculation of the \underline{slope} of experimental
observables in a suitable ``Sudakov regime'', both the constant term
and the unknown mass $M$ would disappear, and only the SUSY parameters
that enter the coefficient of the Sudakov logarithms would be
relevant. In particular, it was shown in~\cite{14} that the process of
top - antitop production would provide an unconventional way for
deriving information, in the MSSM, on the fundamental parameters
$\tan\beta$, in a range of values ($20 < \tan\beta < 40$) that is very
hard to be experimentally explored in other known ways~\cite{15}, in an
energy range $\sqrt{q^2}\simeq 3$ TeV for a reasonable ``light''
(i.e. $< 500$ GeV) $M$ scenario.

The conclusion of the previous analyses is that, in a c.m. energy
range of the TeV size and in a reasonably light $M$ scheme, virtual
SUSY effects would play an important role at the future lepton
colliders in SM pairs production. An almost obvious attitude is, at
this point, that of noticing that, for such $M$ values,
\underline{direct} SUSY pairs production would be copious at such
machines, and of asking whether the SUSY virtual effects for these
processes would be similar, or ``worse'', or ``better''
(i.e. containing more interesting information on parameters) than
those of SM pairs creation.

The aim of this first paper is precisely that of analyzing in this
spirit the process of scalar (i.e. sfermion of Higgs) SUSY pair
production at
lepton colliders, and of showing in some detail which relevant
information on the MSSM parameters would be obtainable from their
virtual effects, and in which possible energy range. Although the
final conclusion are supposed to be valid for the special MSSM case,
an analysis will be performed in a rather general way, so that our
conclusions might be easily generalized for a different type of SUSY
model. In this paper, we will try to be as self contained as possible
working in a first stage at a ``physical'' one-loop level, then
performing a resummation at sub-leading logarithmic order in an
\underline{asymptotic} regime, showing that the two expansions
\underline{do} coincide at one loop. Having proved this equivalence, we
shall perform our numerical analyses in the two different formulations
to show the relevance of a logarithmic resummation in the
``Sudakov regime'' and, in particular, in an energy regime that is
within the reach of the (hopefully near) future lepton colliders, with
a special emphasis on a possible determination of the fundamental MSSM
parameter $\tan\beta$.

Technically speaking, this paper is organized as follows: Section~2 is
devoted to a kinematical description of the processes to be
considered, to a definition of their Born observables and to a
calculation of their expressions at the \underline{one loop} level,
computed in the asymptotic Sudakov regime, In Section~3, the
sub-leading logarithmic order resummation is presented and a comparison
with the one-loop expressions of Section~2 is performed. Section~4
contains a discussion of the size of the virtual Sudakov numerical
effects, both at one loop and resummed, in a ``light $M$'' assumption
for a large class of processes. In Section~5 the special role of the
$\tan\beta$ dependent Yukawa terms in the MSSM is established and the
information derivable on $\tan\beta$ from a measurement of the slopes
of certain \underline{special} processes is exhibited. A final
discussion in Section~6 will then conclude the paper.

\section {SUSY scalar production at the one-loop level}

The aim of this Section is that of giving a general description of SUSY
scalar pair production at the one loop level, at c.m. energies
sufficiently high to justify the use of a logarithmic Sudakov expansion
to describe the leading electroweak behavior of the experimental
observables. This energy range will be denoted, in our pragmatic
definition, as the "electroweak Sudakov regime" of the considered
process.\\
As a first process to be examined, we shall consider that of production
of a (charged \underline{and} neutral) sfermion-antisfermion pair; the
treatment of Higgs pair production will then be derivable with simple
and straightforward modifications. The considered sfermions will be
labelled by their chirality and denoted as $\tilde{f}_{L,R}$. All
sleptons and squarks will be considered, with the exception of
selectrons. For the latter ones, the theoretical description would be
slightly more involved, due to the presence of an extra t-channel
exchange, and we shall postpone it to a next dedicated paper
\cite{nextp}.\\
The results obtained for $\tilde{f}_{L,R}$ production will then
be easily extended to the case of charged or neutral Higgses.

\underline{a) Born level}

\begin{figure}[htb]
\vspace*{1cm}
\[
\epsfig{file=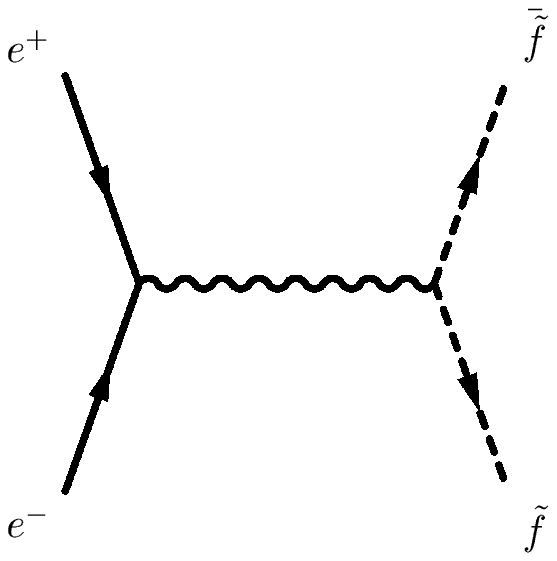,height=4cm}
\]
\vspace*{-0.5cm}
\caption[1]{Born diagram for $e^+e^-\to \bar{\tilde{f}}\tilde{f}$}
\label{diagrama}
\end{figure}

At the Born level, the process is represented by the photon and $Z$
exchange, depicted in Fig.~(\ref{diagrama}). The corresponding decomposition
of the scattering amplitude will be:

\bq
iA^{Born}=i~[~ A^{\gamma,Born}+A^{Z,Born}]
\eq
\noindent
with
\bqa
A^{\gamma,Born}&=&-~{8\pi\alpha_0
Q_f\over q^2}~ \bar v(e^+)~ \gamma^{\mu}
p_{\mu}~u(e^-)
\eqa

\bqa
A^{Z,Born}&=&{4\pi\alpha_0
[I^3_f-s^2_WQ_f]\over s^2_Wc^2_W(q^2-M^2_{Z,0})}~
\bar v(e^+)~ \gamma^{\mu}p_{\mu}~[g^Z_{eL,0}P_L+g^Z_{eR,0}P_R]~ u(e^-)
\eqa
\noindent
where $p$ is the outgoing sfermion momentum, $p'$ the
outgoing anti-sfermion momentum, $q=p+p'$, 
$P_{L,R}=(1\mp\gamma^5)/2$, $\alpha_0$ is the bare QED coupling 
$\alpha_0 = e_0^2/4\pi$, 
$s^2_W$ is the bare Salam-Weinberg angle,
$g^Z_{eL,0}=2s^2_W-1$, $g^Z_{eR,0}=2s^2_W$, $I^3_f$ is the isospin
third component of the final sfermion and $Q_f$ is the sfermion
electric charge in units of $|e|$.

It is rather convenient to introduce the chiral variables $a_{L,R}$.
At the Born level, they are defined as follows:

\bq
A^{Born}\equiv {8\pi\alpha_0\over q^2}~
\bar v(e^+)~\gamma^{\mu} p_{\mu}~ [a^{Born}_L P_L+a^{Born}_R P_R]~u(e^-)
\label{Aborn}\eq

\bq
a^{Born}_L=-Q_f+{(I^3_f-s^2_WQ_f)g^Z_{eL,0}\over2s^2_Wc^2_W}=
-~{s^2_W Q_f+(1-2s^2_W)I^3_f\over2s^2_Wc^2_W}
\label{aLborn}\eq
\bq
a^{Born}_R=-Q_f+{(I^3_f-s^2_WQ_f)g^Z_{eR,0}\over2s^2_Wc^2_W}=
{I^3_f-Q_f\over c^2_W}
\label{aRborn}\eq

We shall first treat, as we said, the processes of production of chiral
sfermion-antisfermion pairs. The meaningful observables that we shall
consider in this paper will be :\\

1) the production cross sections

\bq
\sigma_{L,R}(\tilde{f})=\int_{-1}^1~
d\cos\theta~ [\frac{d\sigma_{L,R}(\tilde{f})}{d\cos\theta}]
\eq

with

\bq
{d\sigma_{L,R}(\tilde{f})\over
dcos\theta}=N{\pi\alpha^2_0\beta^3\over8q^2}~sin^2\theta~
|a_{L,R}(\tilde{f})|^2
\eq
\noindent
where $N$ is the number of colours and $\beta^2=1-4m^2_{\tilde{f}}/q^2$.

At Born level it writes

\bq
\sigma^{Born}_{L,R}(\tilde{f})=N_{col}
{\pi\alpha^2_0\beta^3\over6q^2}~|a^{Born}_{L,R}(\tilde{f})|^2
\eq

Note that, at the Born level, all forward-backward asymmetries
defined with
\bq
\sigma_{L,R}^{FB}(\tilde{f}) = \left(\int_0^1-\int_{-1}^0\right)
d\cos\theta~ [\frac{d\sigma_{L,R}(\tilde{f})}{d\cos\theta}]
\eq

vanish

\bq
A^{Born}_{FB;~L,R}(\tilde{f}) = 
\frac{\sigma^{FB,Born}_{L,R}(\tilde{f})}
{\sigma^{Born}_{L,R}(\tilde{f})}
\equiv 0
\eq
This is a well-known feature of the coupling of a vector boson
to a pair of spinless particles, which cannot generate an
asymmetric $cos\theta$ term.\par
At higher perturbative orders, the forward-backward asymmetries will
not vanish in general. This fact will be stressed and exploited in the
following part of the paper.\\

\vspace{2cm}

2) the various
longitudinal polarization asymmetries

\bq
A_{LR, \tilde{f}} = \frac{\sigma_L(\tilde{f})-\sigma_R(\tilde{f})}
{\sigma_L(\tilde{f})+\sigma_R(\tilde{f})}
\eq

with their Born expressions

\bq
A^{Born}_{LR, \tilde{f}} = 
\frac{|a^{Born}_{L}(\tilde{f})|^2-|a^{Born}_{R}(\tilde{f})|^2}
{|a^{Born}_{L}(\tilde{f})|^2+|a^{Born}_{R}(\tilde{f})|^2}
\eq

\underline{b) General one-loop treatment}\\

In the previous Born expressions, all the involved parameters, i.e. the
electric charge, the $Z$ mass and the Salam-Weinberg angle, are by
definition bare ones, without unambiguously defined physical meanings.
Moving to the next perturbative one-loop level, this ambiguity must be
removed. In the SM case, the general procedures are well known. 
Briefly, the bare charge and the $Z$ mass
are normally replaced by the corresponding \underline{physical}
quantities, defined by measurements performed at $q^2=0$ ("photon
peak") and  at $q^2=M^2_Z$ ("$Z$ peak"). For the bare Salam-Weinberg
angle $s^2_W$, one possible convenient attitude is that of replacing it
by the corresponding effective squared sine $s^2_l$, defined
by measurements at the $Z$ peak, and for a thorough discussion of the
related definitions we defer the reader to the existing literature
\cite{Zsub}. With the \underline{three} previous replacements and
definitions, all the ultraviolet divergences at the one-loop level are
automatically canceled, and the expressions of the various observables
at one-loop may be written as a simple generalization of the
corresponding Born quantities, by formally replacing bare parameters
with physical ones, where special finite gauge-invariant combinations
of one-loop diagrams enter \cite{Zsub}.\\

\begin{figure}[htb]
\vspace*{0.5cm}
\[
\epsfig{file=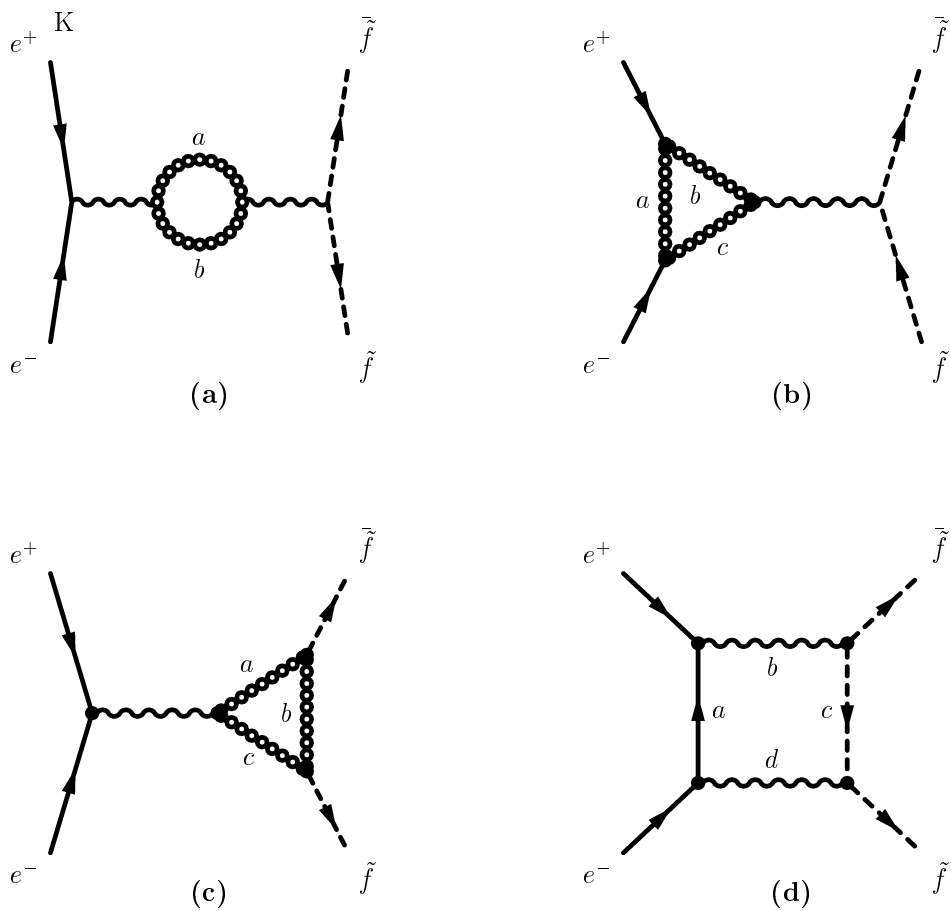,height=12cm}
\]
\vspace*{-0.5cm}
\caption[1]{Typical one loop diagrams
for $e^+e^-\to \bar{\tilde{f}}\tilde{f}$}
\label{diagramb}
\end{figure}

For the processes that we want to consider in this paper, the possible
approach to be followed in the MSSM is essentially similar. This can be
understood without long detailed proofs if one accepts the
prescriptions of the theoretical model, where by definition SUSY is
broken in a "soft" way, so that no new types of ultraviolet divergences
appear with respect to the SM. In particular, for the process that we
are considering, the number of bare parameters that appear at Born
level is still equal to three, so that the definition of three physical
quantities must be sufficient to eliminate all ultraviolet divergences.
Two of these quantities remain the electric charge and the $Z$ mass.
The third bare parameter can still be taken as $s^2_W$, that appears
both in the $Z$ coupling to the initial electron and in that to the
final sfermions. For this quantity, we could always assume a 
redefinition
that implies an extra measurement at a suitable c.m. squared energy
$q^2_{\tilde{f}}$ and "shift" the $q^2_{\tilde{f}}$  dependence in a
one-loop expression, where it will fix e.g. the kinematical point where
to compute finite, gauge-independent combinations of self-energies,
vertices and boxes to be added to a redefined "Born" term now fixed
by $s^2_l(q^2_{\tilde{f}})$. Alternatively, one can still perform 
a redefinition at the "Z  peak" and start from
a Born term that only contains the weak Salam-Weinberg 
angle measured at LEP1, SLC. The price to pay
will be that in the one-loop corrections a fraction of 
the terms will contain contributions to be
theoretically estimated at the correspondent "Z peak" c.m. energy
where experimental information on the final $\tilde{f}\bar{\tilde{f}}$
state does not exist.  
The point is that, in the asymptotic regime in which
we are interested, the dependence on these terms will be 
a part of an overall constant that in the logarithmic content
will disappear. Thus, in practice, the same input parameter 
$s^2_l(M^2_Z)$
that entered the SM case can be taken as the third theoretical input.
This, we stress, will be perfectly acceptable to the extent that one is
only interested in the determination of the leading logarithmic terms
in a high energy expansion, which is exactly our case.
With our choice, the physical expressions that will appear at one loop
will contain, in their so defined "physical" Born approximation, the
same expressions that were entering at the original Born level, with
the bare parameters $\alpha_0$, $M_{0Z}$, $s^2_W$ systematically
replaced by $\alpha$, $M_Z$, $s^2_l(M^2_Z)$ and extra "corrections"
generated by self-energies, vertices and boxes as depicted in 
Fig.~(\ref{diagramb}).
These will bring contributions in the asymptotic energy region
that might, or might not, generate asymptotic logarithms to the various
observables, in a way that we shall now illustrate.\\

\underline{c) Asymptotic behavior of the different one-loop diagrams}\\

The class of diagrams that will generally contribute at one loop is
shown in Fig.2 . To be more precise, we should add the statement that,
in the two diagrams (b), (c) of vertex type, also the various external
self-energy insertions must be included. This will be fundamental in
our approach, since the addition of those diagrams will cancel
ultraviolet divergences of the "normal" vertices, making the overall
contributions ultraviolet finite.\\

At asymptotic energies, the role of the various Figures becomes
drastically different. From Fig.2a we shall obtain the known
Renormalization Group (RG) linear logarithms. These will be generated
from both SM and MSSM virtual pairs. Inside the SM component, there
will be a gauge-dependent term due to virtual pairs of charged $W$'s
and charged would-be SM Goldstone bosons, 
that must be retained in a general
$\xi$-gauge ($\xi\neq \infty$) (our calculations will be
systematically performed in the Feynman-t'Hooft $\xi=1$ gauge, and thus
all the SM would-be Goldstone bosons
contributions must be computed). This
gauge-dependence will be canceled, in a by now well-known way
\cite{DS}, by a component (the "pinch" component \cite{pinch}) of the
corresponding SM vertices, and we do not insist here on this fact, that
has already been exhaustively discussed in previous references
\cite{3}. In our notation, the RG contribution will thus indicate the
subleading (linear) logarithm generated at one loop by the sum of the
self-energies and of the "pinch" components of the vertices.\par
In Fig.2b, the "non pinch" SM component of the initial
vertices must be selected, together with the genuinely supersymmetric
one of the MSSM. This operation will lead to two separate classes of
contributions; the first one, coming from the vertices with SM virtual
exchanges ($~(abc)\equiv (\gamma e e),~(Z e e),~(W\nu\nu),~(\nu WW)~$) 
will generate both quadratic (DL) and
linear (SL) Sudakov logarithms (note that the SM would-be Goldstone
bosons and all Higgs
contributions vanish due to the (assumed) vanishing electron mass).
The genuine SUSY vertices, corresponding to Fig.2b with  
($~(abc)\equiv(\chi^0_i\tilde{e}\tilde{e}),~
(\chi^+_i\tilde{\nu}\tilde{\nu}),~(\tilde{\nu}\chi^+_i\chi^+_j)~$) 
(where we denote by $\chi^0_i$ and $\chi^+_i$ neutralinos and 
charginos and we assume, again, a vanishing electron mass), 
will "only" generate a linear (SL) Sudakov logarithm,
essentially of \underline{gauge} (i.e. \underline{not} of Yukawa)
origin.\par
A similar picture is valid for the final vertices, represented in
Fig.2c (SM virtual exchanges ($~(abc)\equiv
(\tilde{f}\gamma\tilde{f}),~(\tilde{f}Z\tilde{f}),
~(\tilde{f'}W\tilde{f'})~(W\tilde{f'}W)~$)
and genuine SUSY contributions ($~(abc)\equiv
(\tilde{f}\chi^0_i\tilde{f}),~(\tilde{f'}\chi^+_i\tilde{f'}),~
(\chi^0_i\tilde{f}\chi^0_j),~(\chi^+_i\tilde{f'}\chi^+_j)~$). Note that
the diagrams with Higgs exchanges are now present, but vanish
asymptotically and thus disappear in our analysis.
This time, contributions of Yukawa type (only arising from the
Higgsino component of the chargino and neutralino couplings)
must be retained for
\underline{third} family final pairs, and we shall list them in our
formulae with a proper notation. Again, SM diagrams will generate both
DL and SL terms, while SUSY contributions will systematically be of SL
logarithmic gauge type (and for the third family, also of Yukawa
type).\par
Finally, there will be diagrams of box type, represented in Fig.2d.
This time, a welcome simplification will appear, since only SM virtual
gauge $s$-channel exchanges ($~(abcd)\equiv 
(eZ\tilde{f}Z),~(e\gamma\tilde{f}\gamma),~(eZ\tilde{f}\gamma),
~(e\gamma\tilde{f}Z),~(\nu W\tilde{f'}W)~$)
\underline{do} produce asymptotic Sudakov
logarithms (all other SM boxes \underline{vanish} asymptotically),
while \underline{all} SUSY virtual box exchanges have the typical
property of \underline{vanishing} asymptotically. This is a simple
consequence of the spin structure of the corresponding diagrams, that
allows to avoid to perform several involved calculations in the final
asymptotic numerical analysis. The surviving SM diagrams generate two
kinds of contributions. One is of universal DL kind, when half
of it is combined
with a part of the final $WW$ vertex to produce a "universal" 
($4\ln q^2-\ln^2q^2$) term and the other half is combined with a part
of the initial $WW$ vertex to produce a "universal" 
$(3\ln q^2-\ln^2q^2)~$ term. The other one is
of \underline{non universal}
kind, of SL origin and depending on $cos\theta$, where $\theta$ is the
c.m. scattering angle. All these results are essentially similar to the
ones that were found in the case of SM final fermion
pairs production \cite{13}, with the
expected difference that a new $\simeq ~(4\ln q^2-\ln^2q^2)$ 
universal term appears, associated to
the final \underline{scalar} SUSY pair that is produced.\\

After this first qualitative discussion, that we hope to have presented
in a short and understandable way, we are now ready to write the
various logarithmic contributions to the considered process generated
by the relevant one-loop diagrams. We shall divide them in subsets,
that correspond essentially to the four components of Fig.2 (keeping in
mind the previous remarks on vertex "pinch" components), trying to
separate within each subset the specific constituents of different
diagrams. The procedure will first list the effects on the two
independent quantities $a_{L,R}$, defined by
Eq. (\ref{Aborn}-\ref{aRborn}), the effects on the various observables
will follow. Since the list of equations will be rather long, we shall
try to eliminate, as much as possible, definitions and conventions. 
To render a check of our results realistically possible for the
interested reader, we specify here that our SUSY Feynman rules have
been taken from Rosiek's paper \cite{Rosiek}.\\

\underline{d) Logarithmic expansion at one loop of 
the scattering amplitude}\\

At one loop, with our choice of physical inputs, we shall write the
invariant scattering amplitude, for c.m. energy values $q^2\gg M^2_W$, in
the following form:

\bq
A=A^{Born}+A^{1~loop}\equiv {4\pi\alpha\over q^2}~
\bar v(e^+)\gamma^{\mu} p_{\mu}~ [a_LP_L+a_RP_R]~u(e^-)
\eq
with
\bq
a_L=a^{Born}_L +a^{1~loop}_L
\eq
\bq
a_R=a^{Born}_R +a^{1~loop}_R
\eq

Note that we now write the Born terms as
\bq
a^{Born}_L=-Q_f+{(I^3_f-s^2_lQ_f)g_{eL}
\over2s^2_lc^2_l}=
-~{s^2_l(M^2_Z) Q_f+(1-2s^2_l)I^3_f
\over2s^2_lc^2_l}
\eq
\bq
a^{Born}_R=-Q_f+{(I^3_f-s^2_lQ_f)g_{eR}
\over2s^2_lc^2_l}=
{I^3_f-Q_f\over c^2_l}
\eq

with $g_{eL}=2s^2_l-1,~g_{eR}=2s^2_l$.\\

{\bf Complete asymptotic 1-loop results}\\

We shall now write the separate contributions of RG and Sudakov kind.
For the Sudakov terms, we shall divide the initial vertex contributions
from the final ones and the SM virtual effects ("gauge") from the SUSY
ones. Box contributions will be, in our separation, grouped into the
"final gauge" component eqs.(\ref{aLRG}),(\ref{aRRG}).

\bqa
a^{1~loop}_{L,R}&=& a^{RG}_{L,R}~
+a^{in,gauge}_{L,R}~
+a^{in,SUSY}_{L,R}\nonumber\\
&&
+a^{fin,gauge}_{L,R}~
+a^{fin,SUSY}_{L,R}
\label{a1l}\eqa
\noindent

{\bf RG terms}

\bqa
a^{RG}_{L}&=&[~{-\alpha Q_f\over4\pi}\{({32\over9}N-7)^{SM}
+(3+{16N\over9})^{SUSY}\}\nonumber\\
&&+{\alpha\over2\pi s^2_lc^2_l}[(2I^3_f+Q_f(1-4s^2_l)]\{
{1\over3}[{10-16c^2_l\over6}N+{1+42c^2_l\over8}]^{SM}\nonumber\\
&&
-{1\over4}[{13-18s^2_l\over6}+(3-8s^2_l){2N\over9}]^{SUSY}\}
\nonumber\\
&&+~{(I^3_f-s^2_lQ_f)(2s^2_l-1)\alpha\over8\pi s^4_lc^4_l}\{
[{20-40c^2_l+32c^4_l\over9}N+{1-2c^2_l-42c^4_l\over6}]^{SM}
\nonumber\\
&&+[{13-26s^2_l+18s^4_l\over6}
+(3-6s^2_l+8s^4_l){2N\over9}]^{SUSY}\}~]~\ln({q^2\over \mu^2})
\label{aLRG}\eqa

\bqa
a^{RG}_{R}&=&[~{-\alpha Q_f\over4\pi}\{({32\over9}N-7)^{SM}
+(3+{16N\over9})^{SUSY}\}\nonumber\\
&&+{\alpha\over\pi s^2_lc^2_l}[(I^3_f-2s^2_lQ_f]\{
{1\over3}[{10-16c^2_l\over6}N+{1+42c^2_l\over8}]^{SM}\nonumber\\
&&
-{1\over4}[{13-18s^2_l\over6}+(3-8s^2_l){2N\over9}]^{SUSY}\}
\nonumber\\
&&+~{(I^3_f-s^2_lQ_f)\alpha\over4\pi s^2_lc^4_l}\{
[{20-40c^2_l+32c^4_l\over9}N+{1-2c^2_l-42c^4_l\over6}]^{SM}
\nonumber\\
&&+[{13-26s^2_l+18s^4_l\over6}
+(3-6s^2_l+8s^4_l){2N\over9}]^{SUSY}\}~]~\ln({q^2\over \mu^2})
\label{aRRG}
\eqa

The above one-loop logarithms are of course reproduced by simply
inserting in the Born expression for the cross section the running 
couplings $g$ and $g'$ of $SU(2)\times U(1)$ whose scale dependence
is predicted by the MSSM $\beta$ functions, see Sec.(\ref{sec:resummation}).

The parameter $\mu^2$ which appear in the previous equations will be
fixed at $\mu^2=M^2_Z$, which is a natural and consistent choice in our
approach. For further use we shall define the coefficients 
$c^{RG}_{L,R}$ writing:

\bqa
a^{RG}_{L,R}&\equiv&[a^{Born}_{L,R}~{\alpha\over\pi}]~
c^{RG}_{L,R}
\label{cRG}\eqa\\

{\bf initial gauge terms}

\bqa
a^{in,gauge}_{L}&=&[a^{Born}_{L}{\alpha\over16\pi s^2_lc^2_l}]~
\{(1-2s^2_l)^2~[3\ln {q^2\over M^2_Z}-\ln^2{q^2\over M^2_Z}]\nonumber\\
&&
+4s^2_Wc^2_W~[3\ln {q^2\over M^2_\gamma}-\ln^2{q^2\over M^2_\gamma}]
+2c^2_W~[3\ln {q^2\over M^2_W}-\ln^2{q^2\over M^2_W}]\}
\label{aigl}\eqa

\bqa
a^{in,gauge}_{R}&=&[a^{Born}_{R}{\alpha\over16\pi s^2_lc^2_l}]~
\{4s^4_l[3\ln {q^2\over M^2_Z}-\ln^2{q^2\over M^2_Z}]\nonumber\\
&&
+4s^2_lc^2_l~[3\ln {q^2\over M^2_\gamma}-\ln^2{q^2\over M^2_\gamma}]\}
\label{aigr}\eqa

In the above equation, the mass $M_{\gamma}$ refers to the cut-off
which separates in the photon exchange contribution the ultraviolet
from the infrared part. Since the main purpose of this paper is the
determination of asymptotic electroweak effects (neglecting in
particular soft photon emission effects which are determined by QED
only if the experimental cut $\Delta E$ is smaller then
$M_Z \simeq M_W$, we shall set from now on 
$M_{\gamma} = M_Z = M_W$. With
this choice 
we can check that for both the left and right terms we can write

\bqa
a^{in,gauge}_{L,R}&\equiv&[a^{Born}_{L,R}~{\alpha\over\pi}]~
c^{in,gauge}_{L,R}
\label{cing}\eqa
\noindent
with

\bq
c^{in, gauge}_{L,R} = 
\frac{1}{4} \left(
\frac{I_e(I_e+1)}{s_l^2}+\frac{Y_e^2}{4c_l^2} 
\right)[3\ln {q^2\over M^2_W}-\ln^2{q^2\over M^2_W}]
\label{cingIY}\eq
where $Y_e = 2(Q_e-I^3_e)$ and the quantum numbers required in the 
calculation of $a_L$ ($a_R$) are those of left (right) electrons.\\

\newpage

{\bf initial SUSY terms}\\

In the following equations we have introduced different SUSY scales
$M_{ch}$, $M_{neut}$ for the chargino and neutralino contributions,

\bqa
a^{in,SUSY}_{L}&=&[a^{Born}_{L}{-\alpha\over16\pi s^2_lc^2_l}]~
\{[\ln{q^2\over M^2_{neut}}]+2c^2_l[\ln{q^2\over M^2_{ch}}]\}
\eqa
\bqa
a^{in,SUSY}_{R}&=&[a^{Born}_{R}{-\alpha\over16\pi s^2_lc^2_l}]~
4s^2_l[\ln{q^2\over M^2_{neut}}]
\eqa
\noindent
but in practice we will take them equal to a common SUSY scale
$M_{SUSY}$ and write

\bqa
a^{in,SUSY}_{L,R}&\equiv&[a^{Born}_{L,R}~{\alpha\over\pi}]~
c^{in,SUSY}_{L,R}
\label{cins}\eqa
\noindent
with

\bq
c^{in,SUSY}_{L,R} = -\frac{1}{4} \
\left(
\frac{I_e(I_e+1)}{s_W^2}+\frac{Y_e^2}{4c_l^2} 
\right) [\ln{q^2\over M^2_{SUSY}}]
\label{cinsIY}\eq

{\bf final gauge terms}

\bqa
a^{fin,gauge}_{L}&=&{-\alpha\over8\pi s^4_lc^4_l}~\{
[Q_fs^2_l+I^3_f(1-2s^2_W)]~(~(I^3_f-s^2_lQ_f)^2~
[4\ln {q^2\over M^2_Z}-\ln^2{q^2\over M^2_Z}]\nonumber\\
&&+s^2_lc^2_lQ^2_f[4\ln {q^2\over M^2_\gamma}-\ln^2{q^2\over
M^2_\gamma}]~)\nonumber\\
&&+([Q_f{s^2_lc^2_l\over2}-(2I^3_f){c^2_l\over4}]
[4\ln {q^2\over M^2_W}-\ln^2{q^2\over M^2_W}])_{\tilde{f}_L~only}\}
\nonumber\\
&&-~(~{\alpha\over16\pi s^4_l}(2I^3_f)~
[4\ln {q^2\over M^2_{W}}-\ln^2{q^2\over M^2_{W}}]
-{\alpha\over4\pi s^4_l}(2I^3_f)\ln {q^2\over M^2_{W}}
\ln{1+(2I^3_f)cos\theta\over2}~)_{\tilde{f}_L~only}\nonumber\\
&&-{\alpha\over4\pi s^4_lc^4_l}(I^3_f-s^2_lQ_f)^2(2s^2_l-1)^2
\ln {q^2\over M^2_{Z}}
\ln{1-cos\theta\over1+cos\theta}\nonumber\\
&&-{\alpha\over\pi}Q^2_f\ln {q^2\over M^2_{g}}
\ln{1-cos\theta\over1+cos\theta}\nonumber\\
&&+{\alpha\over\pi s^2_lc^2_l}Q_f(I^3_f-s^2_Wq_f)(2s^2_l-1)
\ln {q^2\over M^2_{\gamma Z}}
\ln{1-cos\theta\over1+cos\theta}
\label{afgl}\eqa

\bqa
a^{fin,gauge}_{R}&=&{-\alpha\over4\pi s^2_lc^4_l}~
[Q_f-I^3_f]~\{(I^3_f-s^2_lQ_f)^2~
[4\ln {q^2\over M^2_Z}-\ln^2{q^2\over M^2_Z}]\nonumber\\
&&+s^2_lc^2_lQ^2_f[4\ln {q^2\over M^2_\gamma}-\ln^2{q^2\over
M^2_\gamma}]+({c^2_l\over2}[4\ln {q^2\over M^2_W}-\ln^2{q^2\over
M^2_W}])_{\tilde{f}_L~only}\}\nonumber\\
&&-{\alpha\over\pi c^4_W}(I^3_f-s^2_lQ_f)^2
\ln {q^2\over M^2_{Z}}
\ln{1-cos\theta\over1+cos\theta}\nonumber\\
&&-{\alpha\over\pi}Q^2_f\ln {q^2\over M^2_{g}}
\ln{1-cos\theta\over1+cos\theta}\nonumber\\
&&+{2\alpha\over\pi c^2_l}Q_f(I^3_f-s^2_lQ_f)
\ln {q^2\over M^2_{\gamma Z}}
\ln{1-cos\theta\over1+cos\theta}
\label{afgr}\eqa

Identifying all gauge mass scales with $M_W$ as we did for the
initial gauge terms, one can factorize
the Born amplitudes and write

\bqa
a^{fin,gauge}_{L,R}&\equiv&[a^{Born}_{L,R}~{\alpha\over\pi}]~
c^{fin,gauge}_{L,R}
\label{cfing}\eqa

\bqa
c^{fin,gauge}_L(\widetilde{f}_L) &=&
\frac{1}{4} \left(
\frac{I_f(I_f+1)}{s_l^2}+\frac{Y_f^2}{4c_l^2}
\right)(4\ln {q^2\over M^2_W}-\ln^2{q^2\over M^2_W}) +\nonumber\\
&+& 
\left(\frac{I^3_f}{2s_l^2}
+\frac{Y_f}{4c_l^2}\right)[ln\frac{q^2}{M^2}]ln\frac{1-\cos\theta}
{1+\cos\theta} +\nonumber \\
&+& \frac{1}{4 a_L^{Born}(\widetilde{f}_L)s_l^4} (2I^3_f)
[ln\frac{q^2}{M^2}]ln\frac{1+2I^3_f\cos\theta}{2}
\label{cLLfg}\eqa

\bqa
c^{fin,gauge}_L(\widetilde{f}_R) &=& 
\frac{Y_f^2}{16c_l^2} (4\ln {q^2\over M^2_W}-\ln^2{q^2\over M^2_W}) + \nonumber\\
&+& \frac{Y_f}{2c_l^2}
[ln\frac{q^2}{M^2}]ln\frac{1-\cos\theta}
{1+\cos\theta}
\label{cLRfg}\eqa

\bqa
c^{fin,gauge}_R(\widetilde{f}_L) &=&
\frac{1}{4}\left(
\frac{I_f(I_f+1)}{s_l^2}+\frac{Y_f^2}{4c_l^2}
\right)(4\ln {q^2\over M^2_W}-\ln^2{q^2\over M^2_W}) +\nonumber \\
&+&  
\frac{Y_f}{2c_l^2}[ln\frac{q^2}{M^2}]ln\frac{1-\cos\theta}
{1+\cos\theta}
\label{cRLfg}\eqa

\bqa
c^{fin,gauge}_R(\widetilde{f}_R) &=& 
\frac{Y_f^2}{16 c_l^2} (4\ln {q^2\over M^2_W}-\ln^2{q^2\over M^2_W}) 
+ \nonumber\\
&+& 
\frac{Y_f}{2c_l^2}[ln\frac{q^2}{M^2}]ln\frac{1-\cos\theta}
{1+\cos\theta}
\label{cRRfg}\eqa

\newpage

{\bf final SUSY terms}\\

They are also written in the form

\bqa
a^{fin,SUSY}_{L,R}&\equiv&[a^{Born}_{L,R}~{\alpha\over\pi}]~
c^{fin,SUSY}_{L,R}
\label{cfins}\eqa

with

\bqa
&&c^{fin,SUSY}_L(\tilde{u}_L, \tilde{d}_L) =
c^{fin,SUSY}_R(\tilde{u}_L, \tilde{d}_L)\nonumber\\
&&=\{
-\frac{1}{2}\left(\frac{I_f(I_f+1)}{s_l^2}+\frac{Y_f^2}{4c_l^2}\right) -
\frac{1}{8M^2_Ws^2_l}[{m^2_u\over sin^2\beta}+{m^2_d\over
cos^2\beta}]~\}[\ln{q^2\over M^2_{SUSY}}]
\label{cfsul}\eqa

\bqa
&&c^{fin,SUSY}_L(\tilde{u}_R) = c^{fin,SUSY}_R(\tilde{u}_R)\nonumber\\
&&=
\{-\frac{1}{2}\left(\frac{I_f(I_f+1)}{s_l^2}+\frac{Y_f^2}{4c_l^2}\right)
-\frac{1}{4M_W^2 s_l^2}{m^2_u\over \sin^2\beta}~\}
[\ln{q^2\over M^2_{SUSY}}]
\label{cfsur}\eqa

\bqa
&&c^{fin,SUSY}_L(\tilde{d}_R) = c^{fin,SUSY}_R(\tilde{d}_R) \nonumber\\
&&=
\{-\frac{1}{2}\left(\frac{I_f(I_f+1)}{s_l^2}+\frac{Y_f^2}{4c_l^2}\right)
-\frac{1}{4M_W^2 s_l^2}{m^2_d\over \cos^2\beta}~\}
[\ln{q^2\over M^2_{SUSY}}]
\label{cfsdr}\eqa

\bqa
c^{fin,SUSY}_L(\tilde{l}_L) = c^{fin,SUSY}_R(\tilde{l}_L) =
-\frac{1}{2}\left(\frac{I_f(I_f+1)}{s_l^2}+\frac{Y_f^2}{4c_l^2}\right)
[\ln{q^2\over M^2_{SUSY}}]
\label{cfsll}\eqa

\bqa
c^{fin,SUSY}_L(\tilde{l}_R) = c^{fin,SUSY}_R(\tilde{l}_R) =
-\frac{1}{2}\left(\frac{I_f(I_f+1)}{s_l^2}+\frac{Y_f^2}{4c_l^2}\right)
[\ln{q^2\over M^2_{SUSY}}]
\label{cfslr}\eqa

An important feature of our approach is that, in the asymptotic regime, 
due to the unitarity
properties of the chargino and neutralino mixing matrices, the
only remaining parameters are $\tan\beta$ and
the SUSY scale which appears in the logarithmic terms.
The role of $\tan\beta$ will be discussed in detail
in Sect.V .\\

\newpage

\underline{e) Extension of the results to the case of charged or neutral
Higgses}\\

All the results written above for $\tilde{\bar{f}}_{L,R}
\tilde{f}_{L,R}$ directly apply
to the case of charged Higgses $H^+H^-$ or of neutral Higgses
$H^0_a H^0_{b+2}$, where the labels ($a=1,2,~b=1,2$) refer to
\bq
H^0_1 = H^0,\qquad H^0_2 = h^0,\qquad
H^0_3 = A^0,\qquad H^0_4 = \varphi^0
\eq
($\varphi^0$ is the neutral Goldstone boson), 
defining $p$ as the outgoing $H^-$ or $H^0_a$ momentum and $p'$ the
outgoing $H^+$ or $H^0_{b+2}$ momentum, and
\begin{itemize}

\item
for $H^+H^-$ ~~~~~~ $Q_f=-1~~~~~I^3_f=-{1\over2}~~~~~Y_f=-1$

\item
for $H^0{H^0}'$ ~~~~~~ $Q_f=0~~~~~I^3_f={1\over2}~~~~~Y_f=-1$

\end{itemize}

In the case of \underline{charged Higgs bosons}, the Born terms are

\bq
a^{Born}_L(H)=~{1\over4s^2_lc^2_l}
~~~~~~~~~~~~~~~
a^{Born}_R(H)=~{1\over 2c^2_l}
\eq
\noindent
so that all initial gauge, initial SUSY, final gauge one loop terms
can be taken from the sfermion case with the appropriate values
of $I^3_f$ and $Y_f$ given above. The final SUSY and  
heavy fermion terms are :

\bqa
a^{fin,SUSY}_{L,R}&\equiv&[a^{Born}_{L,R}~{\alpha\over\pi}]~
c^{fin,SUSY}_{L,R}
\label{afinsHch}\eqa

with

\bqa
c^{fin,SUSY}_L(H) &=& c^{fin,SUSY}_R(H) =
\{~-\frac{1}{2}\left(\frac{I_H(I_H+1)}{s_l^2}
+\frac{Y_H^2}{4c_l^2}\right)
\nonumber\\
&-&\frac{3}{8s_l^2 M_W^2} (m_d^2 \tan^2\beta + m_u^2\cot^2\beta)~\}
[\ln{q^2\over M^2_{SUSY}}]
\label{cfinsHch}
\eqa
\noindent
(note the color factor $3$ arising from the quark triangle loop, which
will enhance these contributions as compared to the case of final
squarks).

In the case of \underline{neutral Higgs bosons}
the Born $\gamma$
exchange is missing and the $Z$ exchange is multiplied (with respect to
the previous cases) by $iA_{ab}$ where
\bq
A_{ab} = \left(\begin{array}{cc}
-\sin(\alpha-\beta) & -\cos(\alpha-\beta) \\
-\cos(\alpha-\beta) & \sin(\alpha-\beta)
\end{array}\right)
\eq
\noindent
The specific cases of use in this paper are

\bqa
e^+e^-\to HA &:& A_{ab} = -\sin(\alpha-\beta) \\
e^+e^-\to hA &:& A_{ab} = -\cos(\alpha-\beta)
\eqa
\noindent
with the corresponding Born terms

\bq
a^{L, Born}_{ab} = \frac{1-2s_l^2}{4s_l^2 c_l^2} A_{ab}
~~~~~~~~~~~~
a^{R, Born}_{ab}= -\frac{1}{2c_l^2} A_{ab}
\eq

The initial gauge and initial SUSY one loop terms
can also be taken from the sfermion case with the appropriate values
of $I^3_f$ and $Y_f$ given above.
The final gauge terms are

\bqa
a^{fin,gauge}_{L,R;(ab)}&\equiv&[a^{L,R;Born}_{(ab)}~{\alpha\over\pi}]~
c^{fin,gauge}_{L,R}(ab)
\label{cfingn}\eqa
\noindent
with

\bqa
c^{fin,gauge}_{L}(ab) &=&
\frac{1}{4} \left(
\frac{I_f(I_f+1)}{s_l^2}+\frac{Y_f^2}{4c_l^2}
\right)(4\ln {q^2\over M^2_W}-\ln^2{q^2\over M^2_W}) \nonumber \\
&-& 
\frac{1}{2 a_{L;(ab)}^{Born}(\widetilde{f}_L)s_l^4} (2I^3_f)
[ln\frac{q^2}{M^2}][ln\frac{1-\cos\theta}{2}+
ln\frac{1+\cos\theta}{2}]
\label{cLLfgn}\eqa

\bqa
c^{fin,gauge}_{R}(ab) &=&
\frac{1}{4}\left(
\frac{I_f(I_f+1)}{s_l^2}+\frac{Y_f^2}{4c_l^2}
\right)(4\ln {q^2\over M^2_W}-\ln^2{q^2\over M^2_W})
\label{cRRfgn}\eqa
\noindent
and the final SUSY and heavy fermion terms can also be written as

\bqa
a^{fin,SUSY}_{L,R;(ab)}&\equiv&[a^{L,R;Born}_{(ab)}~{\alpha\over\pi}]~
c^{fin,SUSY}_{L,R}(ab)
\label{afinsHn}\eqa
\noindent
with

\bqa
c^{fin,SUSY}_L(ab) &=& c^{fin,SUSY}_R(ab) = \{-~\frac{1}{2}\left(
\frac{I_H(I_H+1)}{s_l^2}+\frac{Y_H^2}{4c_l^2}\right)\nonumber\\
&-&
\frac{3}{8s_l^2 M_W^2}~f(ab)~\}[\ln{q^2\over M^2_{SUSY}}]  
\label{cfinsHn}\eqa
\noindent
and

\bq
f(HA)={1\over\sin(\beta-\alpha)}[{\sin\alpha\cos\beta
\over\sin^2\beta}m^2_u-{\cos\alpha\sin\beta
\over\cos^2\beta}m^2_d]
\label{fHA}\eq
\bq
f(hA)=-~{1\over\cos(\beta-\alpha)}[{\cos\alpha\cos\beta
\over\sin^2\beta}m^2_u+{\sin\alpha\sin\beta
\over\cos^2\beta}m^2_d]
\label{fhA}\eq

As one can observe, in the neutral Higgs case there appears one more
SUSY parameter ($\alpha$) than in all other cases, which would require
a separate analysis that we feel is beyond the purposes of this first
paper. We shall treat the neutral Higgs production in more general SUSY
models (also beyond the MSSM) in a dedicated forthcoming paper.

\newpage

\underline{f) Logarithmic expansion of the observable quantities}\\

The logarithmic expansion of the various observables can be now
straightforwardly derived from the eqs.(\ref{cRG}), (\ref{cingIY})
(\ref{cinsIY}), (\ref{cfing}), (\ref{cfins}), (\ref{cfinsHch})
,(\ref{cLLfgn}), (\ref{cRRfgn}, (\ref{cfinsHn}). At one loop, we
obtain the following expressions for the polarized angular
distributions:

\bq
{d\sigma_{L,R;f}\over
dcos\theta}=N_{col}{\pi\alpha^2\beta^3\over8q^2}~sin^2\theta~
|a_{L,R;f}|^2
\label{dsig}\eq
\noindent
with the first order expansion

\bq
|a_{L,R;f}|^2=|a^{Born}_{L,R;f}|^2~[1+2\delta_{L,R}(f)]
\label{atot}\eq
\noindent
where

\bqa
\delta_{L,R}(f)&=&[{\alpha\over\pi}]~\{~c^{RG}_{L,R}(f)
+c^{in~gauge}_{L,R}(f)
+c^{fin~gauge}_{L,R}(f)
\nonumber\\
&&
+c^{in~SUSY}_{L,R}(f)
+c^{fin~SUSY}_{L,R}(f)~\}
\label{delta}\eqa 
and the various $c^i_{L,R}(f)$ can be read off the 
corresponding previous eqs.(\ref{cRG}), (\ref{cingIY})
(\ref{cinsIY}), (\ref{cfing}), (\ref{cfins}), (\ref{cfinsHch})
,(\ref{cLLfgn}), (\ref{cRRfgn}, (\ref{cfinsHn}).\\

We can then consider the integrated observables. With this purpose, we
define the Born quantities\footnote{Notice that
$\rho_{L,f_R} = 1/5$ and $\rho_{R, f_R} = 4/5$.}
\bq
\rho_{L, f}={|a^{Born}_L|^2\over|a^{Born}_L|^2+|a^{Born}_R|^2}=
\frac{[Q_f s_l^2 +I_{3,f}(1-2s_l^2)]^2}
{[Q_f s_l^2 +I_{3,f}(1-2s_l^2)]^2+4s_l^4(I_{3,f}-Q_f)^2}
\label{rhol}\eq
\bq
\rho_{R,f}={|a^{Born}_R|^2\over|a^{Born}_L|^2+|a^{Born}_R|^2}=
\frac{4s_l^4(I_{3,f}-Q_f)^2}
{[Q_f s_l^2 +I_{3,f}(1-2s_l^2)]^2+4s_l^4(I_{3,f}-Q_f)^2}
\label{rhor}\eq
and the integration of the relative 1 loop effects with respect
to the scattering angle (note that there are angular dependent parts
in the $\delta_{L,R}(f)$ due to box contributions)

\bq
N^{\rm full}_{L,R;f} = \int_{-1}^1 d\cos\theta \sin^2\theta
~~[\delta_{L,R}(f)] 
\label{nfull}\eq
\bq
N^{\rm FB}_{L,R;f} = \left(\int_{0}^1 -\int_{-1}^0\right)
d\cos\theta \sin^2\theta~~[\delta_{L,R}(f)] 
\label{nfb}\eq

For real $a_{L, R}$ the relative effect in the total cross section
and the absolute shifts in the asymmetries are
\bq
\frac{\delta\sigma}{\sigma} = 2\frac{3}{4}
[\rho_{L,f} N^{\rm full}_{L,f} + \rho_{R,f} N^{\rm full}_{R,f} ]
\label{dsigsig}\eq
\bq
\delta A_{LR, f} = 4\rho_{L,f}\rho_{R,f}\ \frac{3}{4} (N^{\rm full}_{L,f} - 
N^{\rm full}_{R,f})
\label{dalr}\eq
\bq
A_{FB, f} \equiv \delta A_{FB,f}=
{3\over2} ( \rho_{L,f} N^{FB}_{L,f} + \rho_{Rf} N^{FB}_{Rf} )
\label{dafb}\eq

We stress that there is no forward-backward asymmetry at
Born level for a pair of scalar particles. The contribution
from angular independent terms 
cancel, thus, this asymmetry arises at 1-loop order 
and is only due to the
angular dependent box terms appearing in the above
contributions called "final gauge". 
The $W$ box only contributes final scalar doublets ($\tilde f_L$
or charged, neutral Higgses) labelled with
"$\tilde f_L\  only$". The $\gamma+Z$ contribute
all final states. A simple analytical expression results:

\bq
\frac 3 2 N^{FB}_{L,f} = -~\frac\alpha\pi\ (1-4\ln 2)\left\{
\left[ \frac{1}{8s_l^4}\frac{1}{a_{L, f}^{Born}}
\ln\frac{q^2}{M_W^2}\right]_{\tilde f_L\  only} +a^{Born}_{L,f}
\ln\frac{q^2}{M_Z^2}\right\}
\label{nfbl}\eq
\bq
\frac 3 2 N^{FB}_{R,f} = -\frac\alpha\pi\ (1-4\ln 2)
a^{Born}_{R,f}
\ln\frac{q^2}{M_Z^2}
\label{nfbr}\eq
\bq\eq
where we used $M_{\gamma} = M_{\gamma Z} = M_Z$ in the $\gamma+Z$ boxes.

Now we conclude this (unavoidly) long Sect.II, where the complete DL and
SL terms have been computed at the one-loop level. In the forthcoming
Sect.III we shall perform a resummation of all the logarithmic terms
to \underline{subleading} logarithmic accuracy in a convenient
asymptotic (Sudakov) regime, and show that the calculation
\underline{does} coincide at the one-loop level with that of Sect.II.

\newpage

\section{Resummation of subleading Sudakov logarithms in the MSSM}
\label{sec:resummation}

In the previous section we have calculated one loop
Sudakov logarithms in scalar production
at a linear $e^+e^-$ collider at large energies in the MSSM.

With the expected experimental precision in the one percent regime at such a machine,
the need for a theoretical treatment to the same accuracy was already discussed in the
introduction. Recently, the treatment of electroweak Sudakov logarithms in the SM revealed
the fact that for that purpose at least a two loop treatment to SL accuracy
is indicated for energies
above TeV energies \cite{4}. For SM processes a general method of 
obtaining DL corrections to all orders was presented in Ref. \cite{4} (Fadin et al.)
in the context of
the infrared evolution equation method.
If we assume that the mass scale of the sfermions or Higgs particles are not much larger
than the weak scale,
these results can be applied straightforwardly to the MSSM
since the gauge couplings are preserved under supersymmetry and no additional spin 1 particles
are exchanged.

In case the superpartner masses are larger than $500$ GeV, additional double logarithms
need to be taken into account in a way outlined in Ref. \cite{4} (Fadin et al.). 
In the following we assume that we can neglect such
terms, i.e. that all MSSM scalars have a mass below $500$ GeV.

At the subleading level, the situation in general is less 
clear at higher orders.
For SL angular dependent terms, the same reasoning as above goes through since they
originate only from the exchange of spin 1 gauge bosons and can thus be resummed
as in the SM \cite{5}. 
Box-type diagrams exchanging supersymmetric particles in the s-channel do not contribute
to SL angular terms.
The same holds for all universal 
SL corrections which involve the exchange of SM particles \cite{6} since they are
properties of the external particles only. 

\begin{figure}[htb]
\vspace*{2cm}
\[
\epsfig{file=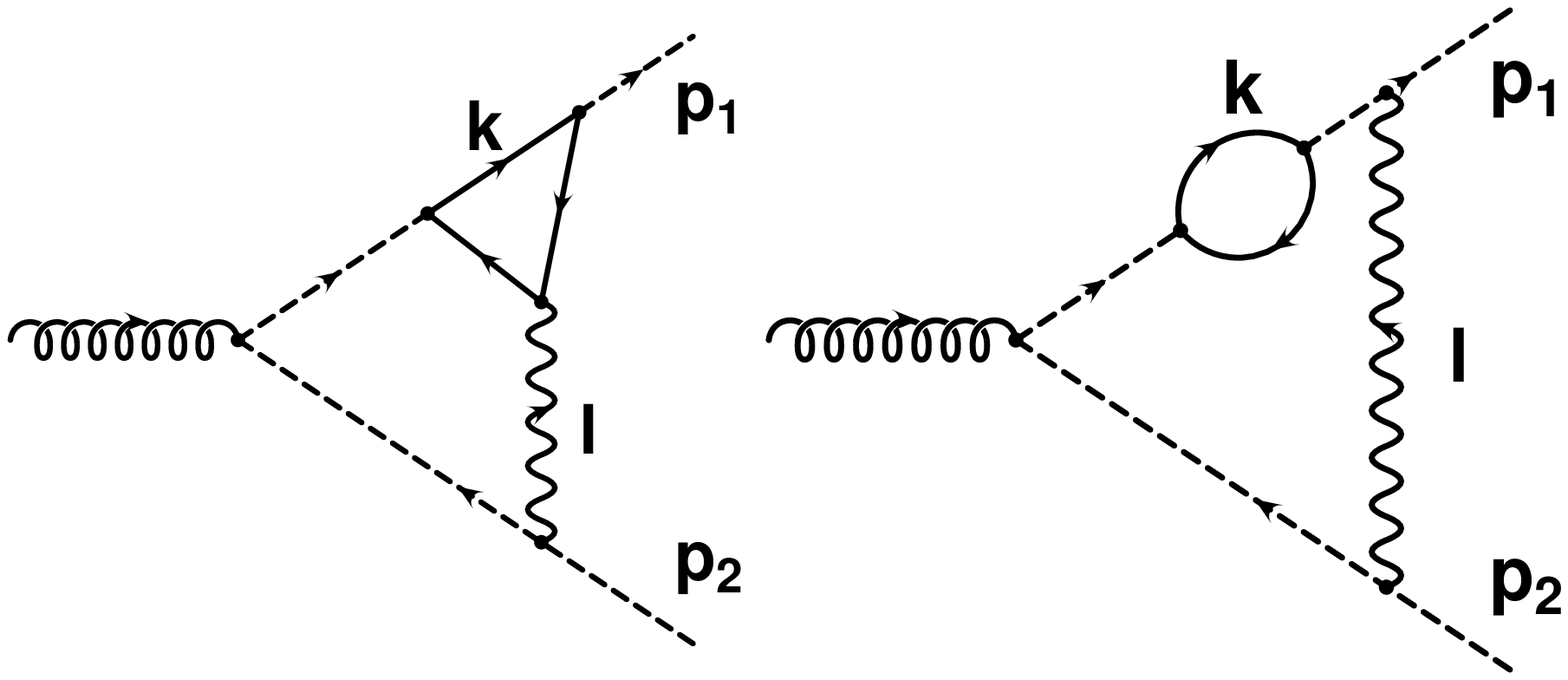,height=6cm}
\]
\vspace*{1cm}
\caption[1]{Two loop corrections involving Yukawa couplings 
of scalars to fermions.
The Ward identity in Eq. (\ref{eq:wi}) assures that in the Feynman gauge, 
the sum
of both diagrams does not lead to 
additional SL logarithms at the two 
loop level.
Only corrections to the original one loop vertex 
(see section \ref{sec:resummation}) need
to be considered and lead to the exponentiation of Yukawa terms in the 
MSSM to SL accuracy.} 
\label{fig:wi}
\end{figure}

New types of
SL Sudakov corrections are, however, involved in the exchange of supersymmetric particles as 
discussed in the previous sections at the one loop level.
We begin with the corrections
contributing in particular the
Yukawa terms from the final state corrections.
In order to see how these corrections enter into two loop SL calculations, we need to
consider the diagrams displayed in Fig. \ref{fig:wi}. 
The corresponding two loop amplitudes read
\begin{eqnarray}
&&\!\!\!\!\!\!\!\!\!\!\!\!\!\! \int \!\! \frac{d^nl}{(4\pi)^n} \!\! \int \!\! \frac{d^nk}{(4\pi)^n} \frac{
(p_1-p_2)_\nu {\rm Tr} \left[
(G_r \omega_r + G_l \omega_l)( \rlap/ k - \rlap/ p_1 ) 2 \rlap/ p_2 ( \rlap/ k - \rlap/ p_1 
+ \rlap/ l ) (G_r \omega_r + G_l \omega_l)
 \rlap/ k \right]}{(l^2-\lambda^2)(p_2+l)^2(p_1-l)^2k^2(k-p_1)^2(k-p_1+l)^2} \label{eq:ver} \\
&&\!\!\!\!\!\!\!\!\!\!\!\!\!\! \int \!\! \frac{d^nl}{(4\pi)^n} \!\! \int \!\! \frac{d^nk}{(4\pi)^n} \frac{
(p_1-p_2)_\nu {\rm Tr} \left[
(G_r \omega_r + G_l \omega_l) ( \rlap/ k - \rlap/ p_1 + \rlap/ l ) (G_r \omega_r + G_l \omega_l)
 \rlap/ k \right] 4 p_1 p_2}{(l^2-\lambda^2)(p_2+l)^2(p_1-l)^2k^2(k-p_1+l)^2(p_1-l)^2} 
 \label{eq:se}
\end{eqnarray}
where we omit common factors and the scalar masses assuming $m_s \sim \lambda$ for clarity.
The $G_{r,l}$ denote the chiral Yukawa couplings and $\omega_{r,l}=\frac{1}{2} \left(1 \pm 
\gamma_5 \right)$. The gauge coupling is written in the symmetric basis for clarity since
we are considering a regime where $q^2=(p_1-p_2)^2 \gg M^2$, where $M$ is the gauge boson mass. In any
case, local gauge invariance is not violated in the SM and for heavy particles in the
high energy limit, we can
perform the calculation in a basis which is more convenient.
For our purposes we need to
investigate terms containing three large logarithms in those diagrams. Since the fermion
loops at one loop only yield a single logarithm it is clear that the gauge boson loop momentum
$l$ must be soft. Thus we need to show that the UV logarithm originating from the $k$
integration is identical (up to the sign) in both diagrams.
We can therefore neglect the loop momentum $l$ inside the fermion loop.
We find
for the fermion loop vertex $\Lambda^\mu(p_1^2,0,p_1^2)$ belonging to Eq. (\ref{eq:ver}):
\begin{eqnarray}
&& \frac{{\rm Tr} \left[(G_r \omega_r + G_l \omega_l)( \rlap/ k - \rlap/ p_1 ) \gamma^\mu ( \rlap/ k
- \rlap/ p_1 ) (G_r \omega_r + G_l \omega_l)\rlap/ k \right]}{k^2(k-p_1)^2(k-p_1)^2} \nonumber \\
&=& \frac{4G_rG_l \left(2 p_1^\mu (k^2-p_1k) +k^\mu (p_1^2-k^2) \right)}{k^2(k-p_1)^4}
\end{eqnarray}
This we need to compare with the self energy loop $\Sigma (p_1^2)$ from Eq. (\ref{eq:se}):
\begin{eqnarray}
&& \frac{\partial}{\partial {p_1}_\mu} \frac{{\rm Tr} \left[ (G_r \omega_r + G_l \omega_l)
( \rlap/ k - \rlap/ p_1 ) (G_r \omega_r + G_l \omega_l)  \rlap/ k \right]}{
k^2(k-p_1)^2} \nonumber \\
&=& \frac{\partial}{\partial {p_1}_\mu} \frac{4G_rG_l(p_1k-k^2)}{k^2(k-p_1)^2}
= 4 G_rG_l \frac{2p_1^\mu(k^2-p_1k)+k^\mu(p_1^2-k^2)}{k^2(k-p_1)^4}
\end{eqnarray}
In short we can write
\begin{equation}
\frac{\partial}{\partial {p_1}_\mu} \Sigma (p_1^2)= \Lambda^\mu (p_1^2,0,p_1^2)
\label{eq:wi}
\end{equation}
Thus, we have established a Ward identity for arbitrary Yukawa couplings of scalars to
fermions and thus, the identity of the UV singular contributions. 
\underline{The relative sign is
such} \underline{that the generated SL logarithms of the diagrams in Fig. \ref{fig:wi} cancel each
other}. The existence of such an identity is not surprising since it expresses the fact
that also the Yukawa sector is gauge invariant since supersymmetry preserves the gauge symmetry.
in perturbation theory to SL accuracy.
For the same reason the SM-Yukawa terms were found to exponentiate in Refs. \cite{6}.
Also in an axial gauge the corrections can be seen to factorize accordingly since
in this gauge DL terms originate only from on-shell two point functions.

\begin{figure}[htb]
\vspace*{2cm}
\[
\epsfig{file=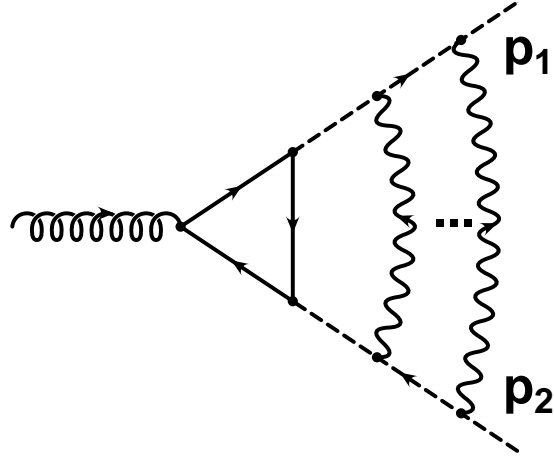,height=6cm}
\]
\vspace*{1cm}
\caption[1]{Higher order corrections to vertices with Yukawa couplings to
SL accuracy. The graph is only schematic since in principle the gauge bosons couple
to all external legs in the process. Due to the discussion in the text the non-Abelian version
of Gribov's factorization theorem can be employed in the context of the
infrared evolution equation method.} 
\label{fig:gt}
\end{figure}

We are thus left with gauge boson corrections to the original vertices in the on-shell
renormalization scheme such as depicted in Fig. \ref{fig:gt}. 
At high energies we can therefore employ the non-Abelian version
of Gribov's bremsstrahlung theorem in accordance with the SM case \cite{4} (Fadin et al.).

\begin{figure}[htb]
\vspace*{2cm}
\[
\epsfig{file=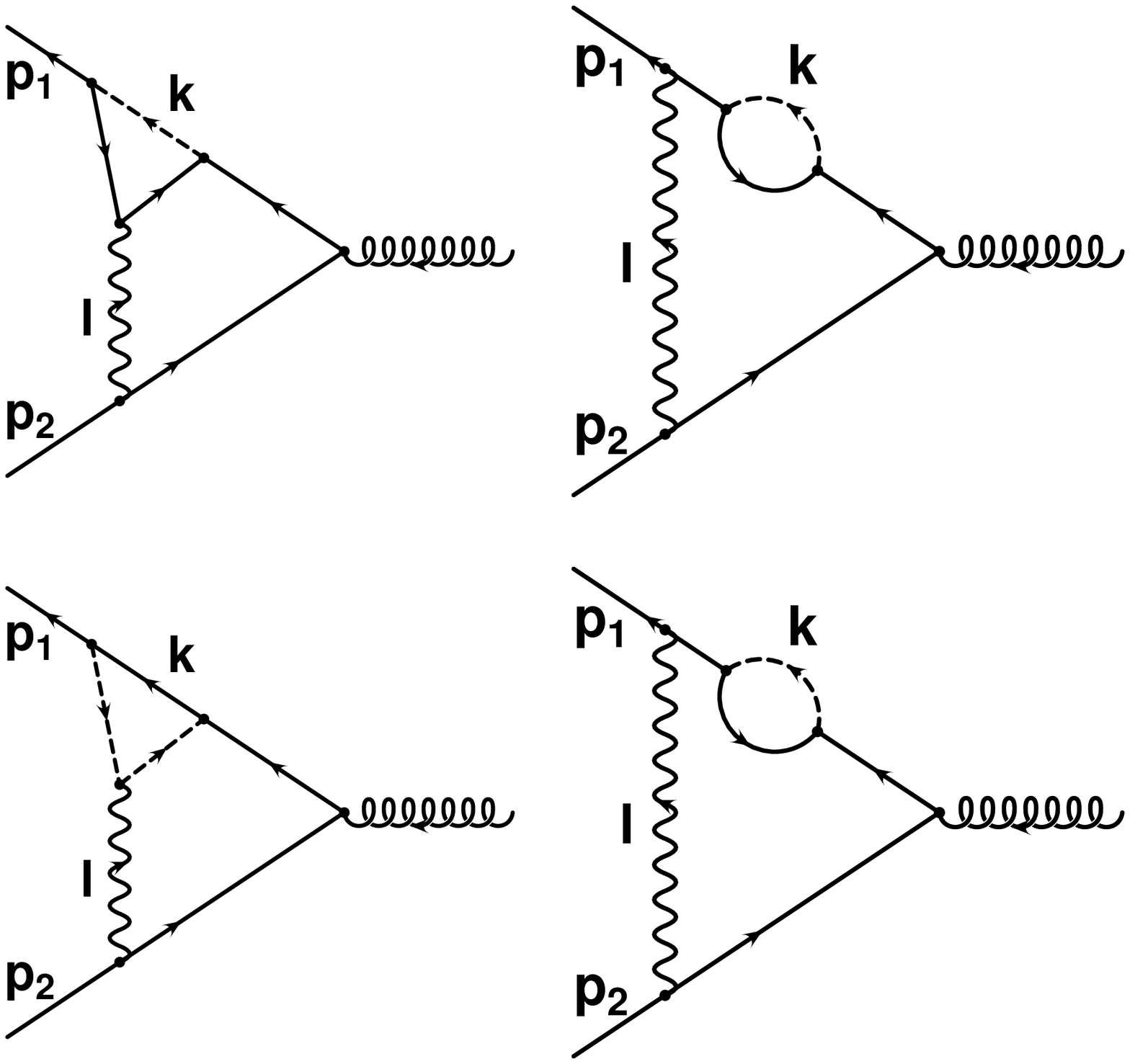,height=12cm}
\]
\vspace*{1cm}
\caption[1]{Two loop corrections involving susy couplings 
of scalars to fermions.
The Ward identity in Eq. (\ref{eq:wi}) assures that in the Feynman gauge, 
the sum
of both diagrams in each row does not lead to 
additional SL logarithms at the two 
loop level.
Only corrections to the original one loop vertex 
(see section \ref{sec:resummation}) need
to be considered and lead to the exponentiation of gauge terms in the 
MSSM to SL accuracy.} 
\label{fig:sgwi}
\end{figure}

Analogously, it is easy to see that also the diagrams depicted in Fig. \ref{fig:sgwi}
form the initial state exchange of supersymmetric scalar particles lead to Ward
identities. For the respective vertex and self energy contributions we have verified
that the corresponding Eq. (\ref{eq:wi}) is fulfilled. Thus the same reasoning as above
can be applied and the exponentiation at the SL level is established.

For our purposes here we omit the soft photon regime for now (which is determined by QED
only if we impose an experimental energy resolution below the weak scale \cite{6}) 
and focus only on the novel higher order Sudakov corrections in the MSSM.
For clarity and later convenience, we use a common mass scale in all logarithms below.
This is not fully correct to SL accuracy in the DL terms as discussed below but
can easily be rectified using the scales found in the one loop calculation presented
in the previous sections.
In the high energy regime one then has the following result for sfermion production
to SL-accuracy
relative to the Born cross section 
\footnote{We denote the chirality L,R by the
the index $\alpha$ with $-L=R$.}
(with $t=-\frac{q^2}{2}(1-\cos \theta)\;,\;\;u=-\frac{q^2}{2}
(1+\cos \theta)$):
\begin{eqnarray}
d \sigma^{\rm SL}_{e^+_{-\alpha} e^-_{\alpha} \longrightarrow {\overline {\tilde f}}_{-\beta} 
{\tilde f}_{\beta}} 
\!\!&=&\!\!\! d \sigma^{\rm Born}_{e^+_{-\alpha} e^-_{\alpha} \longrightarrow {\overline {\tilde f}}_{-\beta}
{\tilde f}_{\beta}} \times \nonumber \\ && \!\!\!\!\!\!\!\!\!\!\!\!\!\!\!\!\!\!\!\!
\exp \left\{ - \frac{g^2(m_{\tilde f}^2)}{4\pi^2} I_{e^-_{\alpha}} \left( I_{e^-_{\alpha}}+1
\right) \left[ \frac{1}{c} \ln \frac{q^2}{m_{\tilde f}^2}
\left( \ln \frac{g^2(m_{\tilde f}^2)}{g^2
(q^2)} - 1 \right) + \frac{1}{c^2}
\ln \frac{g^2(m_{\tilde f}^2)}{g^2(q^2)} \right] \right. \nonumber \\
&&\!\!\!\!\!\!\!\!\!\!\!\!\!\!\!\!\!\!\!\! -\frac{{g^\prime}^2(m_{\tilde f}^2) Y^2_{e^-_{\alpha}}}{16 \pi^2 } 
\left[
\frac{1}{c^\prime} \ln \frac{q^2}{m_{\tilde f}^2}
\left( \ln \frac{{g^\prime}^2(m_{\tilde f}^2)}{{g^\prime}^2
(q^2)} - 1 \right) + \frac{1}{{c^\prime}^2}
\ln \frac{{g^\prime}^2(m_{\tilde f}^2)}{{g^\prime}^2(q^2)} \right] \nonumber \\
&&\!\!\!\!\!\!\!\!\!\!\!\!\!\!\!\!\!\!\!\! + \left( \frac{ g^2(m_{\tilde f}^2)}{8 \pi^2}  
I_{e^-_{\alpha}} \left( I_{e^-_{\alpha}}+1 \right)+
\frac{ {g^\prime}^2(m_{\tilde f}^2)}{8 \pi^2} \frac{Y^2_{e^-_{\alpha}}}{4} \right)  3 \ln \frac{q^2}{m_{\tilde f}^2}
\nonumber \\ &&\!\!\!\!\!\!\!\!\!\!\!\!\!\!\!\!\!\!\!\!  
- \frac{g^2(m_{\tilde f}^2)}{4\pi^2} I_{{\tilde f}_{\beta}} \left( I_{{\tilde f}_{\beta}}+1
\right) \left[ \frac{1}{c} \ln \frac{q^2}{m_{\tilde f}^2}
\left( \ln \frac{g^2(m_{\tilde f}^2)}{g^2
(q^2)} - 1 \right) + \frac{1}{c^2}
\ln \frac{g^2(m_{\tilde f}^2)}{g^2(q^2)} \right] \nonumber \\
\!\!\!\!&&\!\!\!\!\!\!\!\!\!\!\!\!\!\!\!\!\!\!\!\! -\frac{{g^\prime}^2(m_{\tilde f}^2) 
Y^2_{{\tilde f}_{\beta}}}{16 \pi^2 } 
\left[
\frac{1}{c^\prime} \ln \frac{q^2}{m_{\tilde f}^2}
\left( \ln \frac{{g^\prime}^2(m_{\tilde f}^2)}{{g^\prime}^2
(q^2)} - 1 \right) + \frac{1}{{c^\prime}^2}
\ln \frac{{g^\prime}^2(m_{\tilde f}^2)}{{g^\prime}^2(q^2)} \right] \nonumber \\
\!\!\!\!&&\!\!\!\!\!\!\!\!\!\!\!\!\!\!\!\!\!\!\!\! + \left( \frac{ g^2(m_{\tilde f}^2)}{8 \pi^2} 
I_{{\tilde f}_{\beta}} \left( I_{{\tilde f}_{\beta}}+1 \right)+
\frac{ {g^\prime}^2(m_{\tilde f}^2)}{8 \pi^2} \frac{Y^2_{{\tilde f}_{\beta}}}{4} \right)  4 \ln \frac{q^2}{
m_{\tilde f}^2}
+ {\tilde \delta}_{
e^+_{-\alpha} e^-_{\alpha} \longrightarrow {\overline {\tilde f}}_{-\beta} {\tilde f}_{\beta}} \ln
\frac{q^2}{m^2_{\tilde f}}
\nonumber \\ && \!\!\!\!\!\!\!\!\!\!\!\!\!\!\!\!\!\!\!\!
-\frac{g^2 (m_{\tilde f}^2)}{8\pi^2} \ln \frac{q^2}{m_{\tilde f}^2} \left[ \left( \tan^2 \theta_{\rm w} Y_{e^-_{\alpha}} Y_{
{\tilde f}_\beta}
+ 4 T^3_{e^-_{\alpha}} T^3_{{\tilde f}_\beta} \right) \ln \frac{t}{u} \right. \nonumber \\ && \!\!\!
\!\!\!\!\!\!\!\!\!\!\!\!\!\!\!\!\! \left.
\left. + \frac{\delta_{\alpha, {\rm L}} \delta_{\beta,{\rm L}}}{\tan^2 \theta_{\rm w} Y_{e^-_{\alpha}} Y_{
{\tilde f}_\beta} /4
+ T^3_{e^-_{\alpha}} T^3_{{\tilde f}_\beta}} \left( \delta_{d,{\tilde f}} 
\ln \frac{-t}{q^2}
- \delta_{u,
{\tilde f}}
\ln \frac{-u}{q^2} \right)  \right] \right\} \label{eq:sfang}
\end{eqnarray}
where the last line only contributes for left handed (L) electrons
and the $d,u$ symbols denote the
corresponding isospin quantum number of ${\tilde f}$. 
In addition we denote $c=\frac{g^2 (m_{\tilde f}^2) {\tilde \beta}_0}{
4\pi^2}$, $c^\prime=\frac{{g^\prime}^2 
(m_{\tilde f}^2) {\tilde \beta}^\prime_0}{4\pi^2}$, where $g$, $g'$ are
the $SU(2)_l\times U(1)$ gauge couplings.
Here we assume that the asymptotic MSSM $\beta$ functions can be used with
\begin{eqnarray}
{\tilde \beta}_0&=& \frac{3}{4} C_A- \frac{n_g}{2}-\frac{n_h}{8} \;\;,\;
{\tilde \beta}_0^\prime=-\frac{5}{6}n_g-\frac{n_h}{8} \\
g^2 (q^2) &=& \frac{g^2(m_{\tilde f}^2)}{1+{\tilde \beta}_0 \frac{ g^2
(m_{\tilde f}^2)}{4\pi^2}
\ln \frac{q^2}{m_{\tilde f}^2}} \;\;,\;
{g^\prime}^2 (q^2) = \frac{{g^\prime}^2 (m_{\tilde f}^2)}{1+{\tilde \beta}^\prime_0
\frac{{g^\prime}^2 (m_{\tilde f}^2)}{4\pi^2}
\ln \frac{q^2}{m_{\tilde f}^2}}
\end{eqnarray}
where $C_A=2$, $n_g=3$ and $n_h=2$. In practice, one has to use the relevant numbers of active
particles in the loops. These terms correspond to the RG-SL corrections just as in the case
of the SM as discussed in Refs. \cite{6} but now with the MSSM particle spectrum contributing.
They originate only from RG terms within loops which without the RG contribution
would give a DL correction.
It must be mentioned that the one-loop RG corrections do not exponentiate and are omitted
in the above expression! They are, however, completely determined by the renormalization group
in softly broken supersymmetric theories such as the MSSM
and sub-subleading at the higher than one loop order.

The generic term denoted by ${\tilde \delta}_{e^+_{-\alpha} e^-_{\alpha} \longrightarrow 
{\overline {\tilde f}}_{-\beta} {\tilde f}_{\beta}}$ in Eq. (\ref{eq:sfang})
is a short hand notation for the overall (initial and final) 
one loop SUSY corrections discussed in the previous section.
The result in Eq. (\ref{eq:sfang}) is given for the chiral
superpartners directly. In principle, mixing effects need to be taken into 
account for the mass eigenstates of the third family as
discussed in Sect.IV. 

Eq. (\ref{eq:sfang}) contains all SL terms to all 
orders under the assumptions stated above.
In particular it provides an independent check on 
the diagrammatic one-loop results derived
in Sect.2 for all universal corrections from gauge bosons as well as the
angular dependent terms.

As mentioned above, to compare the asymptotic expansion (3.6) (valid in an energy
regime where the details of electroweak symmetry breaking can be
neglected) with a \underline{physical} one-loop calculation like the
one that we performed in Sect.II, a number of "minor" adjustments must
be performed. In practice, one should use, rather than a common mass
$m_{\tilde{f}}$, the gauge boson masses $M_W$, $M_Z$ or the SUSY mass
$M_{SUSY}$ in the corresponding logarithms. 
After these replacements, the one-loop version of Eq. (3.6) 
should reproduce the corresponding results of
Sect.II.\par
We have verified that the asymptotic expansion Eq. (3.6) and the
corresponding physical one-loop expressions of Sect.II do actually
coincide. This can be verified in a reasonably simple way, which we do
not show in detail here to avoid writing extra long equations. The
result can be considered, in our opinion, a satisfactory check of both
the various theoretical arguments presented in this Section and of the detailed
calculations of Sect.II.\\ 

In the case of charged Higgs production we have analogously:
\begin{eqnarray}
d \sigma^{\rm SL}_{e^+_{-\alpha} e^-_{\alpha} \longrightarrow H^+ H^-} 
\!\!&=&\!\!\! d \sigma^{\rm Born}_{e^+_{-\alpha} e^-_{\alpha} \longrightarrow H^+ H^-}
\times \nonumber \\ && \!\!\!\!\!\!\!\!\!\!\!\!\!\!\!\!\!\!\!\!
\exp \left\{ - \frac{g^2(m_H^2)}{4\pi^2} I_{e^-_{\alpha}} \left( I_{e^-_{\alpha}}+1
\right) \left[ \frac{1}{c} \ln \frac{q^2}{m_H^2}
\left( \ln \frac{g^2(m_H^2)}{g^2
(q^2)} - 1 \right) + \frac{1}{c^2}
\ln \frac{g^2(m_H^2)}{g^2(q^2)} \right] \right. \nonumber \\
&&\!\!\!\!\!\!\!\!\!\!\!\!\!\!\!\!\!\!\!\! -\frac{{g^\prime}^2(m_H^2) Y^2_{e^-_{\alpha}}}{16 \pi^2 } 
\left[
\frac{1}{c^\prime} \ln \frac{q^2}{m_H^2}
\left( \ln \frac{{g^\prime}^2(m_H^2)}{{g^\prime}^2
(q^2)} - 1 \right) + \frac{1}{{c^\prime}^2}
\ln \frac{{g^\prime}^2(m_H^2)}{{g^\prime}^2(q^2)} \right] \nonumber \\
&&\!\!\!\!\!\!\!\!\!\!\!\!\!\!\!\!\!\!\!\! + \left( \frac{ g^2(m_H^2)}{8 \pi^2}  
I_{e^-_{\alpha}} \left( I_{e^-_{\alpha}}+1 \right)+
\frac{ {g^\prime}^2(m_H^2)}{8 \pi^2} \frac{Y^2_{e^-_{\alpha}}}{4} \right)  3 \ln \frac{q^2}{m_H^2}
\nonumber \\ &&\!\!\!\!\!\!\!\!\!\!\!\!\!\!\!\!\!\!\!\!  
- \frac{g^2(m_H^2)}{4\pi^2} I_H \left( I_H+1
\right) \left[ \frac{1}{c} \ln \frac{q^2
}{m_H^2}
\left( \ln \frac{g^2(m_H^2)}{g^2
(q^2)} - 1 \right) + \frac{1}{c^2}
\ln \frac{g^2(m_H^2)}{g^2(q^2)} \right] \nonumber \\
\!\!\!\!&&\!\!\!\!\!\!\!\!\!\!\!\!\!\!\!\!\!\!\!\! -\frac{{g^\prime}^2(m_H^2) 
Y^2_H}{16 \pi^2 } 
\left[
\frac{1}{c^\prime} \ln \frac{q^2}{m_H^2}
\left( \ln \frac{{g^\prime}^2(m_H^2)}{{g^\prime}^2
(q^2)} - 1 \right) + \frac{1}{{c^\prime}^2}
\ln \frac{{g^\prime}^2(m_H^2)}{{g^\prime}^2(q^2)} \right] \nonumber \\
\!\!\!\!&&\!\!\!\!\!\!\!\!\!\!\!\!\!\!\!\!\!\!\!\! + \left( \frac{ g^2(m_H^2)}{8 \pi^2}  
I_H \left( I_H+1 \right)+
\frac{ {g^\prime}^2(m_H^2)}{8 \pi^2} \frac{Y^2_H}{4} \right)  4 \ln \frac{q^2}{
m_H^2}
+ {\tilde \delta}_{
e^+_{-\alpha} e^-_{\alpha} \longrightarrow H^+ H^-} \ln
\frac{q^2}{m^2_H}
\nonumber \\ && \!\!\!\!\!\!\!\!\!\!\!\!\!\!\!\!\!\!\!\! \left.
-\frac{g^2 (m_H^2)}{4\pi^2} \ln \frac{q^2}{m_H^2} \left[ 
\delta_{\alpha,{\rm L}} \left( \frac{1}{2 c^2_{\rm w}} \ln \frac{t}{u} + 2 c^2_{\rm w}
\ln \frac{-t}{q^2} \right) 
+ \delta_{\alpha, {\rm R}} \tan^2 \theta_{\rm w}
\ln \frac{t}{u}  \right] \right\} \label{eq:Hang}
\end{eqnarray}
where we denote $c=\frac{g^2(m_H^2) {\tilde \beta}_0}{
4\pi^2}$, $c^\prime=\frac{{g^\prime}^2 (m_H^2) {\tilde \beta}^\prime_0}{4\pi^2}$.

The generic term denoted by ${\tilde \delta}_{e^+_{-\alpha} e^-_{\alpha} \longrightarrow H^+H^-}$
in Eq. (\ref{eq:Hang}) is again a short hand notation for the 
one loop SUSY corrections discussed in the previous section.

We emphasize again the independent nature of deriving the remaining 
results in Eq. (\ref{eq:Hang}) as they are directly
obtained from $e^+e^-\rightarrow \phi^+ \phi^-$ \cite{2,6}
if we neglect the mass-difference.
A comparison with the one loop calculation confirms again the 
corresponding contributions of
the explicit diagrammatic calculation of Sect.II.

Having checked the equality of the asymptotic resummed subleading
expansion and of the physical one loop calculation for all SUSY
production processes (modulo $\tilde{e},~ \tilde{\nu_e}$ to be
considered later on), in the next sections 
we turn to a discussion of the phenomenological importance of
the one loop and resummed corrections particularly
for a determination of $\tan \beta$.

\section{Numerical Results}

Given the fact that we have now at our disposal both a physical
one-loop calculation and a \underline{consistent} SL order resummation,
we are in a position to compute all relevant observables in the two
approximations and to identify, whenever this occurs, the importance of
the extra resummation in the MSSM case.\\

To summarize the numerical weight of the various contributions
to the one loop cross sections 
\bq
e^+e^-\to \widetilde{f}_{L,R}\widetilde{f}_{L,R}^*, H^+H^-
\eq
we approximate the previous results by
replacing in the logarithms $M_W$,  $M_\gamma$ and $M_{\gamma Z}$
with $M_Z$ and by using everywhere a single SUSY mass scale 
$M_{SUSY}$ in place of the various $M_{\chi^+}$, $M_{\chi^0}$ and 
$M_{\widetilde{f}}$. Under this approximation, we obtain
\bqa
100\ \frac{\delta \sigma}{\sigma} &=& (k^{RG}_{SM}+ k^{RG}_{SUSY}) 
\ln\frac{q^2}{\mu^2} + \nonumber \\
&+& k^{in}_{gauge} L_3\left(\frac{q^2}{M_Z^2}\right) + 
k^{in}_{SUSY} \ln\frac{q^2}{M_S^2} + \nonumber \\
&+& k^{fin}_{gauge} L_4\left(\frac{q^2}{M_Z^2}\right) + 
k^{fin}_{SUSY} \ln\frac{q^2}{M_S^2} + \nonumber \\ 
&+& \left(k^{Yuk}_t \frac{m_t^2}{M_W^2\sin^2\beta} +
k^{Yuk}_b \frac{m_b^2}{M_W^2\cos^2\beta}\right) \ln\frac{q^2}{M_S^2} + 
k^{Box} \ln\frac{q^2}{M_Z^2}
\eqa
where
\bq
L_n(x) = n \ln x - \ln^2 x
\eq
and 
$k^{RG}_{SM}$, $k^{RG}_{SUSY}$, $k^{in}_{gauge}$, $k^{in}_{SUSY}$,
$k^{fin}_{gauge}$, 
$k^{fin}_{SUSY}$, $k^{Yuk}_{t,b}$, $k^{Box}$
are numerical coefficients.
Their detailed origin is the following
\begin{enumerate}
\item[--] $k^{RG}_{SM}$, $k^{RG}_{SUSY}$: Renormalization Group
terms, Eqs.(\ref{aLRG},\ref{aRRG}).

\item[--] $k^{in}_{gauge}$, $k^{in}_{SUSY}$: initial gauge and SUSY terms 
according to Eqs.(\ref{cingIY},\ref{cinsIY}).

\item[--] $k^{fin}_{gauge}$: final gauge terms, namely the 
 $\theta$ independent logarithms in
Eqs.(\ref{afgl},\ref{afgr})
with $M_W$ set to $M_Z$.

\item[--] $k^{fin}_{SUSY}$: final SUSY terms; these are the $\tan\beta$
independent terms appearing in Eqs.(\ref{cfsul}-\ref{cfslr}).

\item[--] $k^{Yuk}_{t,b}$: final ``massive'' SUSY terms; these are the 
$\tan\beta$ dependent terms proportional to $1/\sin^2\beta$ and $1/\cos^2\beta$
respectively in Eqs.(\ref{cfsul}-\ref{cfslr}).

\item[--] $k^{Box}$: all the terms resulting from the angular integration
of the $\theta$ dependent terms (note that we are now separating box
terms from final gauge ones, differently from what we did in
eqs.(\ref{afgl})-(\ref{cRRfg}) ).
\end{enumerate}

The following remarkable relations hold
\bq
k^{in}_{gauge} = -k^{in}_{SUSY} \stackrel{\mbox{\tiny def}}{=} k^{in}
\eq
\bq
k^{fin}_{SUSY} = -2 k^{fin}_{gauge} 
\eq
and allow to write the simpler expression
\bqa
\label{kdef}
100\ \frac{\delta \sigma}{\sigma} &=& (k^{RG}_{SM}+ k^{RG}_{SUSY}) 
\ln\frac{q^2}{\mu^2} + \nonumber \\
&+& k^{in}\left[ L_3\left(\frac{q^2}{M_Z^2}\right) -
\ln\frac{q^2}{M_S^2}\right] + 
k^{fin}_{gauge} \left[L_4\left(\frac{q^2}{M_Z^2}\right)
-2 \ln\frac{q^2}{M_S^2} \right]  + \nonumber \\
&+&  
\left(k^{Yuk}_t \frac{m_t^2}{M_W^2\sin^2\beta} +
k^{Yuk}_b \frac{m_b^2}{M_W^2\cos^2\beta}\right) \ln\frac{q^2}{M_S^2} + 
k^{Box} \ln\frac{q^2}{M_Z^2}
\eqa

The numerical values of $k^{RG}_{SM}$, $k^{RG}_{SUSY}$, $k^{in}$, 
$k^{fin}_{gauge}$, $k^{Yuk}_{t,b}$ and  $k^{Box}$
are shown in Tab.~(\ref{tab1}).

In the case of the longitudinal and forward-backward 
asymmetries we compute instead
the absolute percentual effect $100\ \delta A$ and group 
the various contributions
exactly as in the previous equation. 
The numerical values of the coefficients for $A_{LR}$
are also shown in Tab. (\ref{tab1}). In the case of $A_{FB}$, 
only $k^{Box}$ is non zero and its values are 
\bq
A_{FB}:\qquad
k^{Box} = \left\{\begin{array}{ccccc}
\widetilde{t}_L & \widetilde{t}_R & \widetilde{b}_L & 
\widetilde{b}_R & H^\pm \\
-1.3 & -0.32 & 1.3 & 0.16 & 1.1
\end{array}
\right.
\eq

We shall now illustrate the main results of our analysis in the
following figures. We must anticipate at this point that the
qualitative features of the various effects show a strong difference
between the two cases of\\
 a) production of sleptons in general (with the
exclusion of selectrons and $\tilde{\nu_e}$) and squarks of the first
two families and\\
 b) production of squarks of the third family and of
charged Higgs bosons. 
One can already guess that the difference will be due
to the appearance, in the second case, of contributions of SUSY Yukawa
type, that will give rise to relevant and possibly interesting effects
to be discussed later on.\par
\begin{figure}[htb]
\vspace*{2cm}
\[
\psfig{file=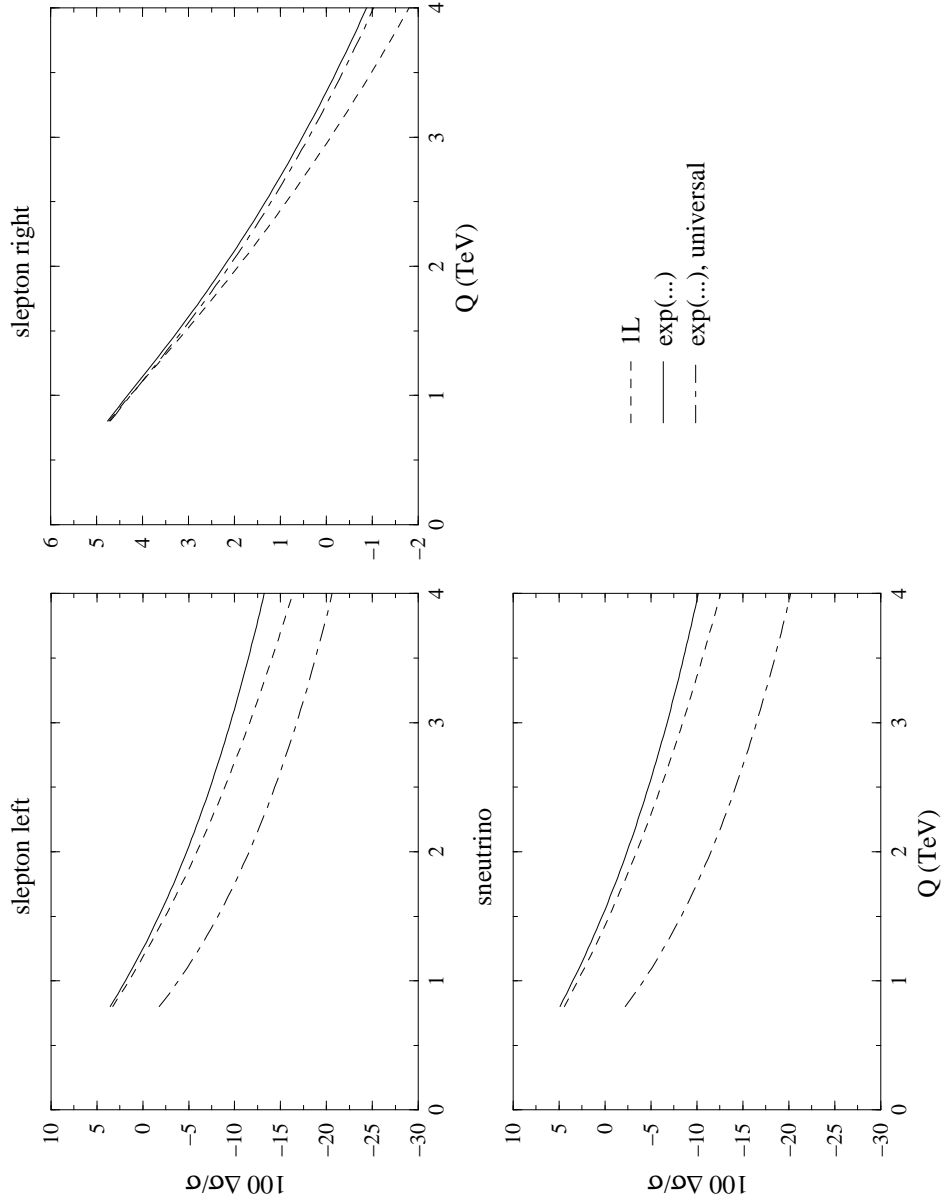,height=16cm}
\]
\vspace*{1cm}
\caption[1]{Cross section for production of sleptons ($\neq \widetilde{e}$)
or sneutrinos.
``1L'' is the total (Renormalization Group and Sudakov) one loop virtual 
effect. ``exp(...)'' is the cross section resummed at subleading order 
including RG contributions. In the last curve, labeled 
``exp(...), universal'', the angular dependent terms coming from boxes
have been suppressed. In this and the following Figures we denote by $Q$
the c.m. energy $\sqrt{q^2}$.}
\label{sigmasll}
\end{figure}
\begin{figure}[htb]
\vspace*{2cm}
\[
\psfig{file=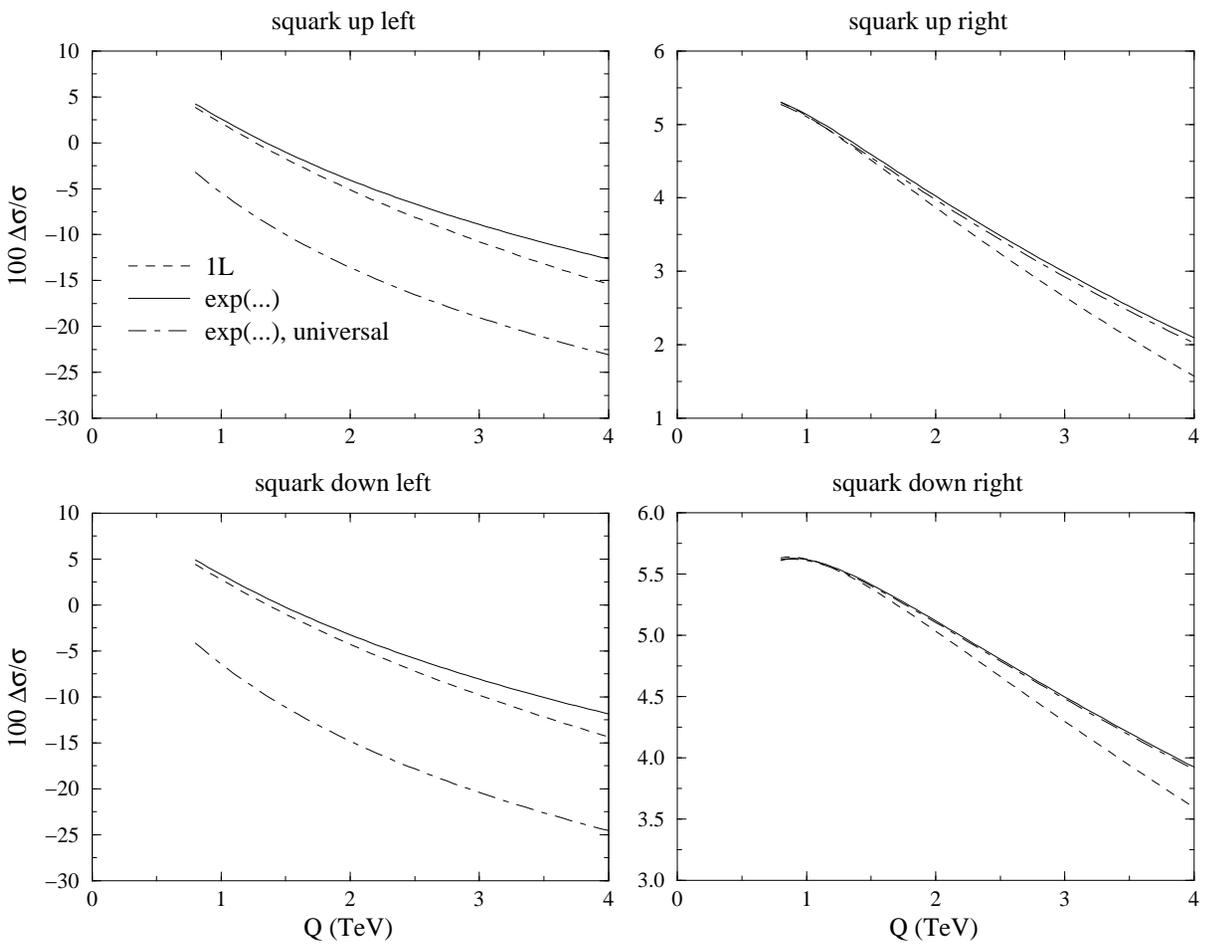,height=16cm}
\]
\vspace*{1cm}
\caption[1]{Cross section for production of up or down squarks
in the first two generations. The labeling of the various curves is
explained in the caption of Fig.~\ref{sigmasll}.}
\label{sigmasquark}
\end{figure}
\begin{figure}[htb]
\vspace*{2cm}
\[
\psfig{file=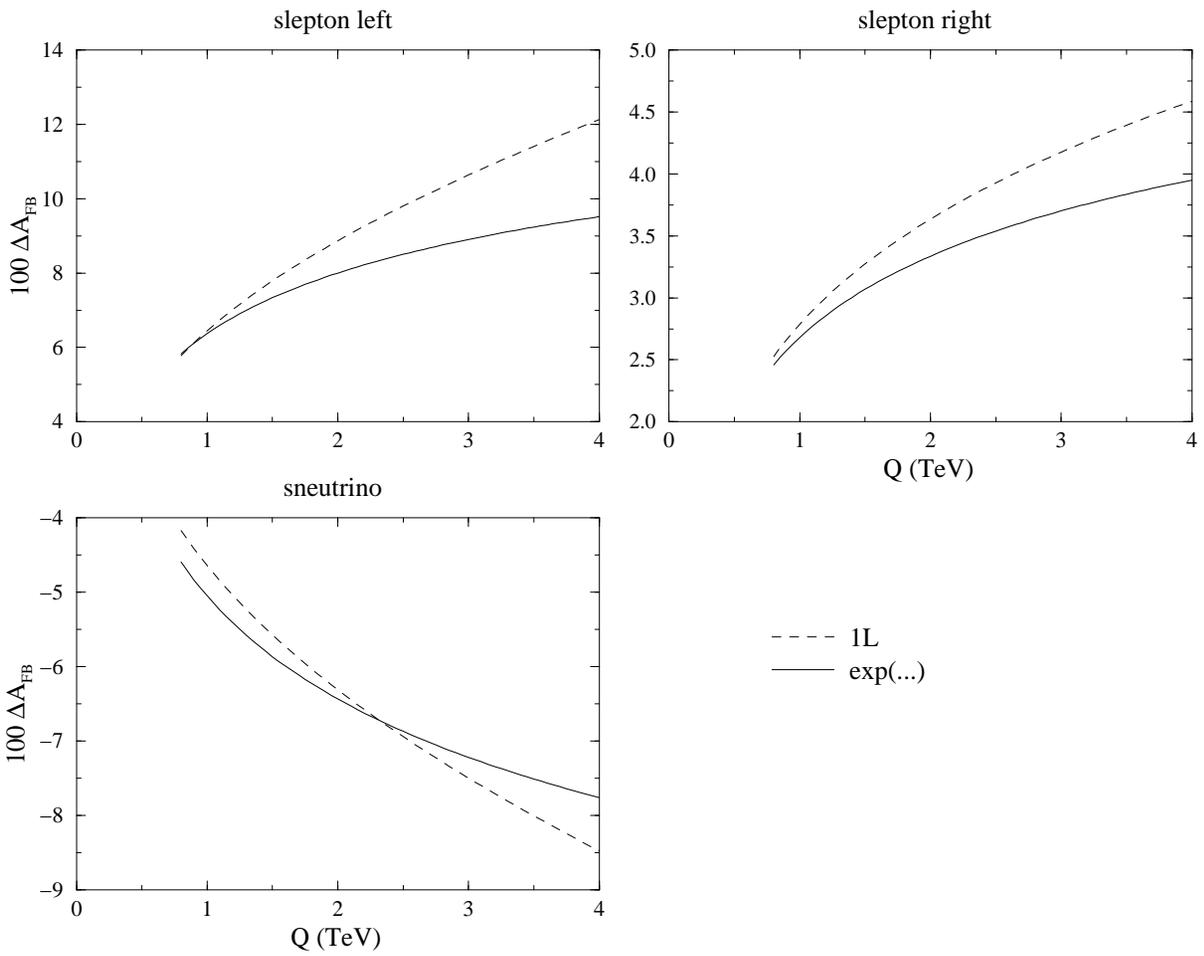,height=16cm}
\]
\vspace*{1cm}
\caption[1]{Forward-Backward asymmetry for production of sleptons 
($\neq \widetilde{e}$) or sneutrinos.
The labeling of the various curves is
explained in the caption of Fig.~\ref{sigmasll}.}
\label{afbsll}
\end{figure}
\begin{figure}[htb]
\vspace*{2cm}
\[
\psfig{file=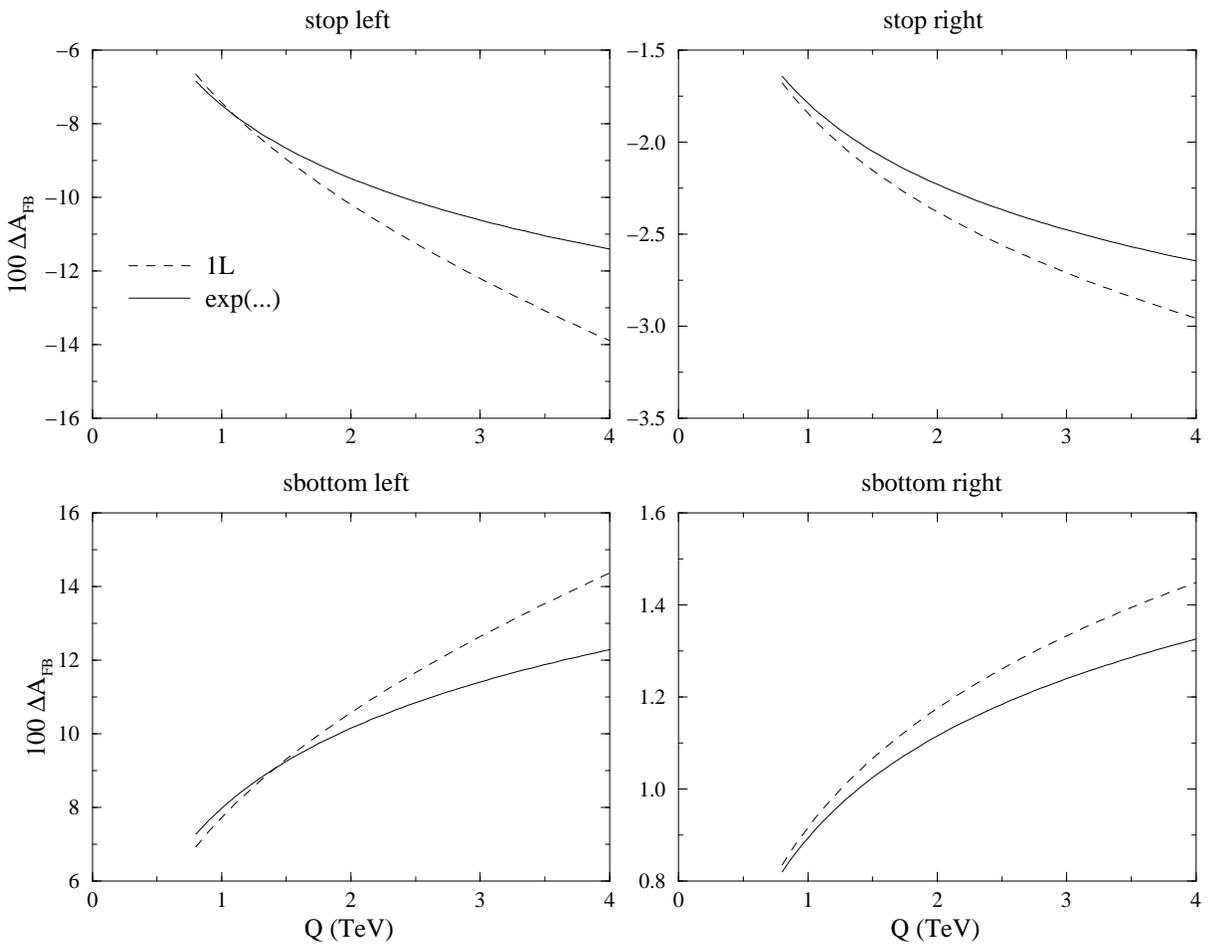,height=16cm}
\]
\vspace*{1cm}
\caption[1]{Forward-Backward asymmetry for production of up or down
squarks in the first two generations.
The labeling of the various curves is
explained in the caption of Fig.~\ref{sigmasll}.}
\label{afbsquark}
\end{figure}

Fig.\ref{sigmasll} shows the relative logarithmic effects 
at one loop (dashed line)
and after resummation (full line) for $\tilde{\mu}_{L,R}$,
$\tilde{\tau}_{L,R}$, $\tilde{\nu}_{\mu,L}$, $\tilde{\nu}_{\tau,L}$,
cross sections at variable energy, while in Fig.\ref{sigmasquark}
 the same effects are
shown for the first two families of squarks (denoted as $\tilde{u},~
\tilde{d}$. Figs.\ref{afbsll},\ref{afbsquark} 
correspond to the cases of the forward-backward
asymmetries of the same processes. In all Figures, the "logarithmic"
effect takes into account both the RG SL and the Sudakov DL,SL
terms.\par
As one sees from Figs.\ref{sigmasll}-\ref{afbsquark}, 
the general feature is that the logarithmic
effect at one loop in the
energy region between 1 TeV and 2 TeV 
is very close to the resummed effect, up to less than one 
percent differences
that in our working assumption should be experimentally invisible. This
feature, that is valid for both cross sections and forward-backward
asymmetries, is intuitively related to the fact that the one-loop
effects are in these cases relatively "small", although experimentally
meaningful, being typically of several percent size. The
same conclusions apply to the cases of the longitudinal polarization
asymmetries of this first class of processes. In order to avoid a too
large number of Figures, we have given the corresponding effects in
Tables 2,3. The situation changes when one moves to larger energies,
$2~TeV\lesssim\sqrt{q^2}\lesssim4~TeV$
(the latter value is chosen as an optimistic 
"averaged" aim for CLIC experiments). Here, for
\underline{left-handed} sleptons and squarks, the difference between
one loop and resummation becomes appreciable (beyond the one percent
level) and a complete resummation appears necessary.\par
The conclusion of this first investigation is that, in the energy range
of the TeV size, assuming a typical SUSY mass of few hundred GeV, the
perturbative treatment of SUSY scalar production for the considered set
of processes appears to us to be totally satisfactory at the one-loop
level, not requiring extra resummations if the aimed experimental
accuracy remains at the one percent (and not at the one permille) level
(we stress that, if the SUSY mass turned out to be larger, these
conclusions can be simply rescaled at correspondingly larger c.m.
energies). For larger energies, as resummation seems to be requested
for left-handed scalars.\par
As a side remark that might be added, we have also depicted in 
Figs.\ref{sigmasll}-\ref{afbsquark}
the separate resummed effects that would have been obtained by ignoring
the SL $\theta$-dependent contributions of box origin. As one sees, the
consequence of this omission would have been catastrophic in the left
sfermion cases, in particular, the angular terms are the
\underline{only ones} that contribute the forward-backward asymmetries.
This confirms a previous observation \cite{8} that stressed the
relevance of these "box-type" contributions in the Sudakov regime.\par
\begin{figure}[htb]
\vspace*{2cm}
\[
\psfig{file=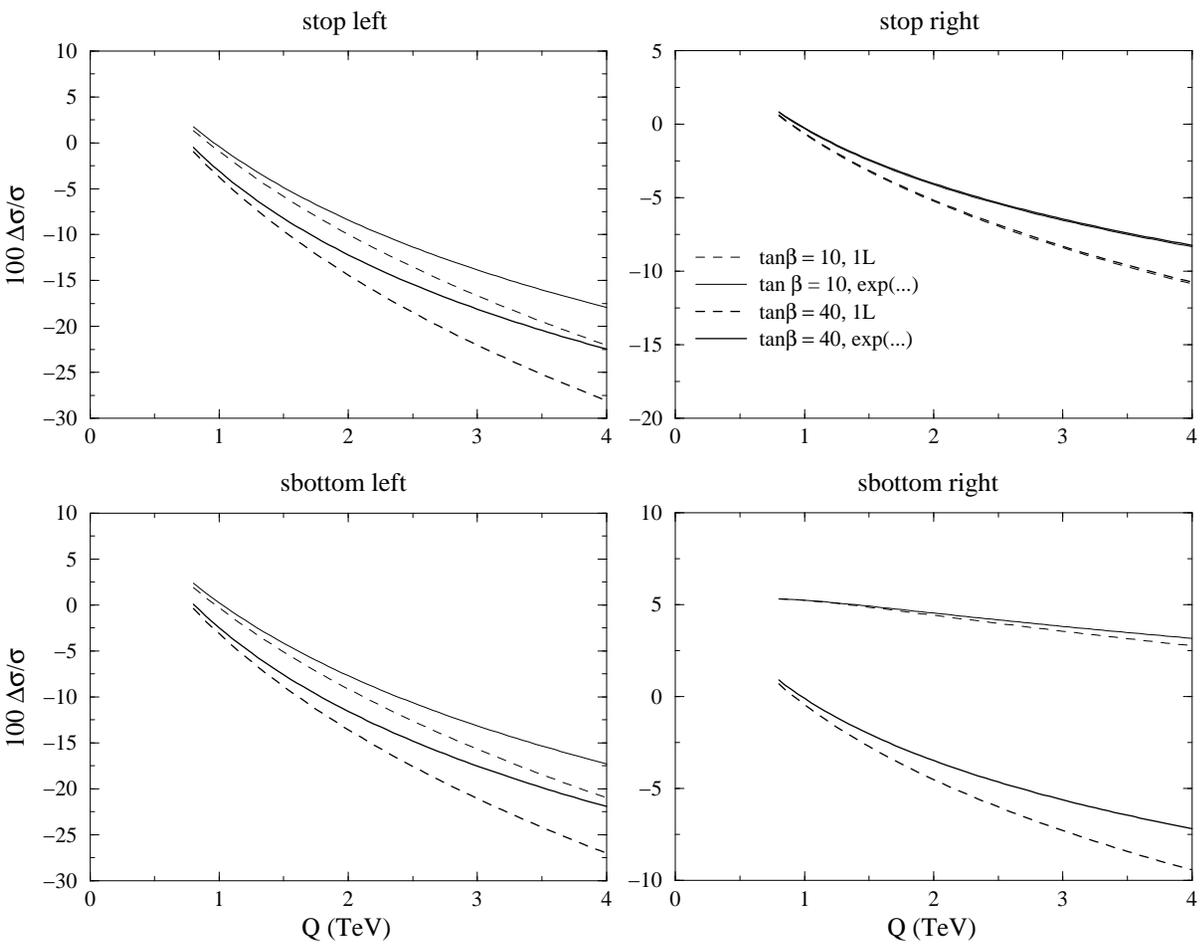,height=16cm}
\]
\vspace*{1cm}
\caption[1]{Cross section for production of third generation
squarks (stop and sbottom). We show the one loop (1L) and resummed
(exp(...)) shifts at $\tan\beta = 10, 40$.}
\label{tanbetastl}
\end{figure}
\begin{figure}[htb]
\vspace*{2cm}
\[
\psfig{file=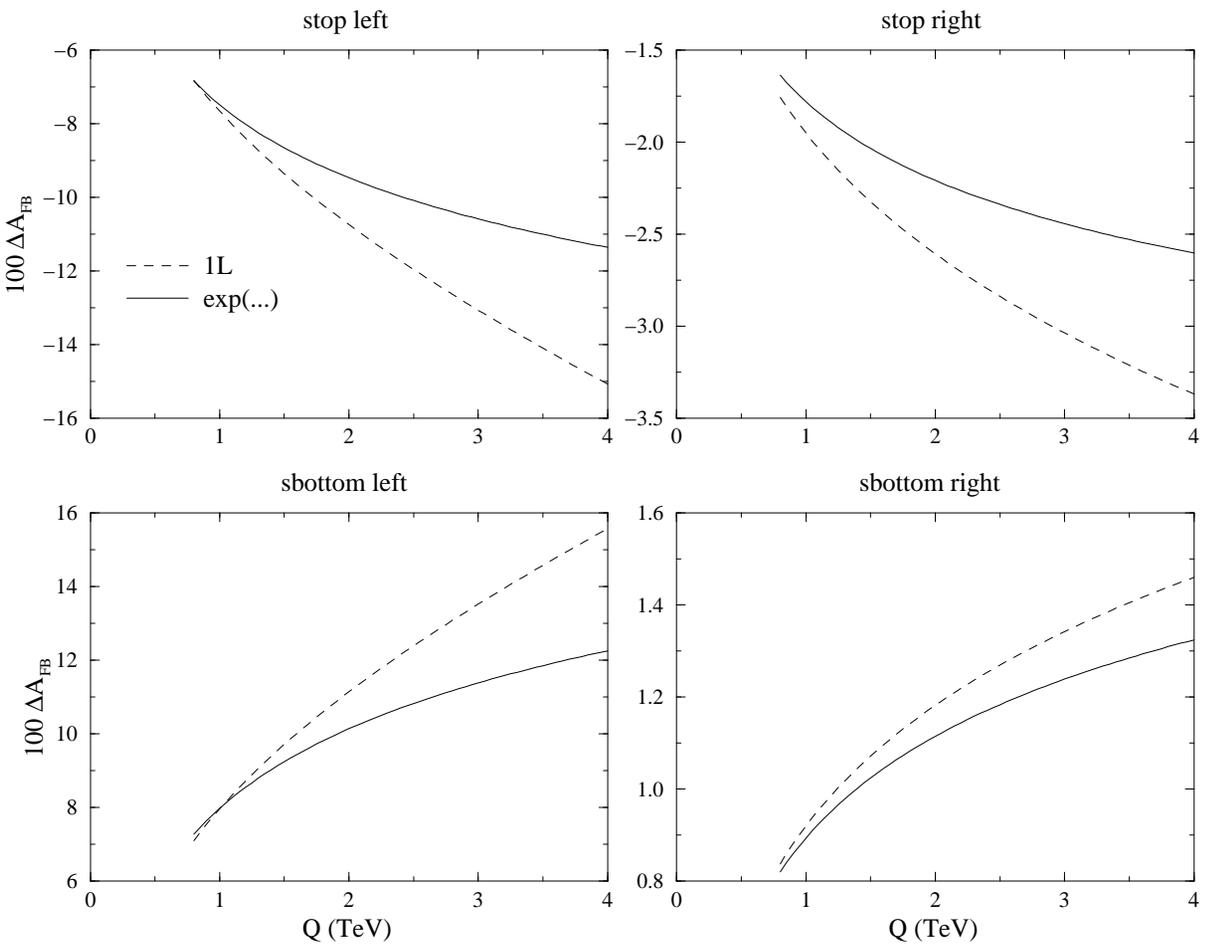,height=16cm}
\]
\vspace*{1cm}
\caption[1]{Forward-Backward asymmetry for production of third generation
squarks (stop and sbottom). We show the one loop (1L) and resummed
(exp(...)) shifts. Here, all the effect is due to the angular dependent terms
coming from boxes and Yukawa terms depending on $\tan\beta$ cancel.}
\label{afbstl}
\end{figure}
\begin{figure}[htb]
\vspace*{2cm}
\[
\psfig{file=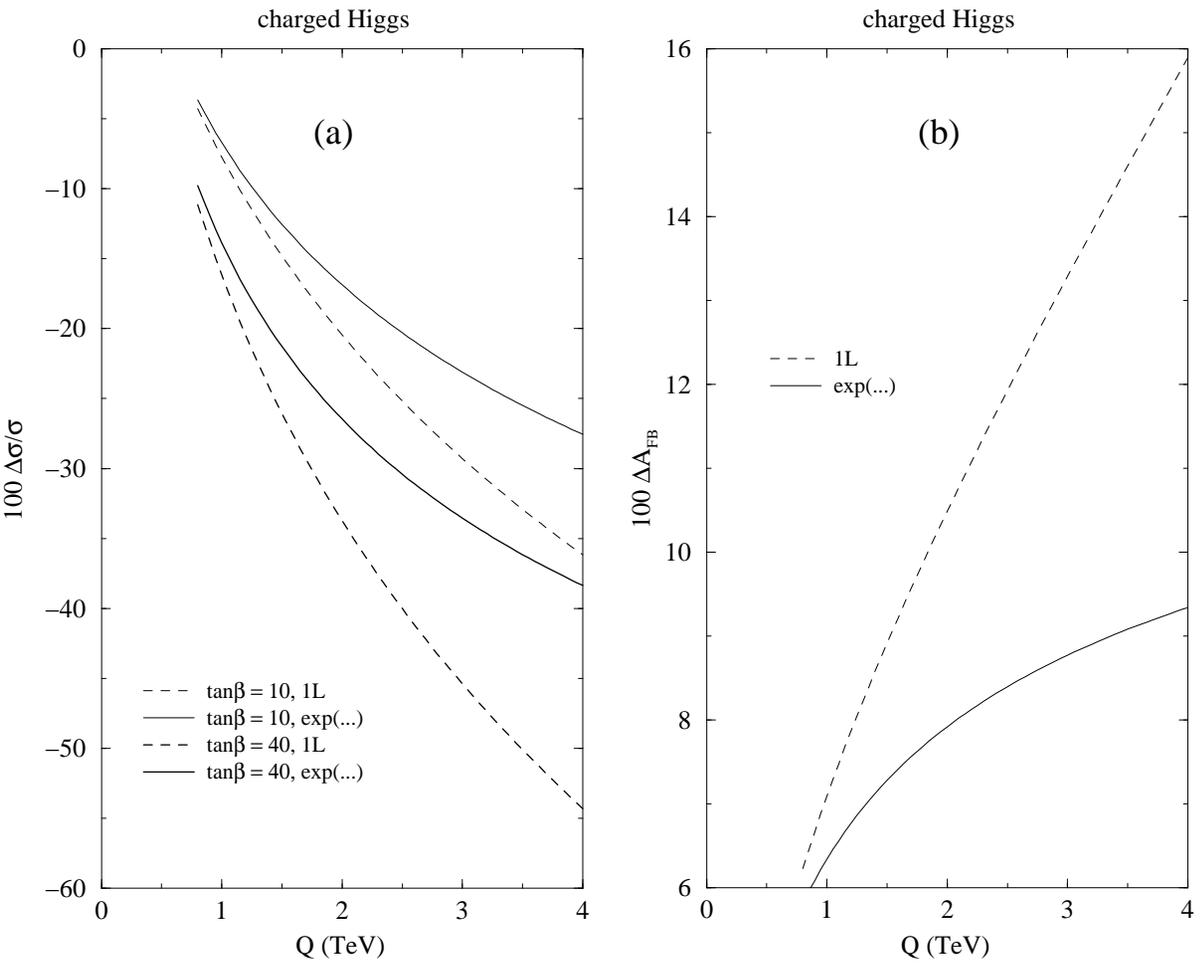,height=16cm}
\]
\vspace*{1cm}
\caption[1]{Cross section (a)
and forward-backward asymmetry (b) for production
of charged Higgs bosons. For both observables 
we show the one loop (1L) and
resummed (exp(...)) shifts at $\tan\beta=10,40$. For the asymmetry we
remind that there are no $\tan\beta$ dependent Yukawa terms.} 
\label{sigmaafbhiggs}
\end{figure}

When we move to the production of either third family of squarks or
charged Higgs bosons (we remind that neutral Higgs bosons production
will be treated in detail in a forthcoming paper), a 
different picture arises. A priori, we know that SUSY Yukawa
contributions will be effective in these cases, and they will depend on
the $\tan\beta$ parameter. This statement is only true for cross
sections, and does not hold for forward-backward asymmetries, since the
latter ones are not affected by universal, $\theta$-independent,
contributions. To evidentiate whether this expectation is correct, 
we have
depicted in Figs.\ref{tanbetastl},\ref{sigmaafbhiggs}a 
the effects on the cross sections for squarks
and Higgs bosons production at two representative and sensibly
different values $\tan\beta=10$ and $\tan\beta=40$. Again, we have drawn
systematically the one loop (dashed) and resummed (full) effects. In
Figs.\ref{afbstl},\ref{sigmaafbhiggs}b  
we have drawn the forward-backward asymmetries of Higgs
bosons and squarks production that, as we said, do not depend on
$\tan\beta$.
The \underline{first} characteristic feature of 
Figs.\ref{tanbetastl},\ref{sigmaafbhiggs}a is that now
a more drastic difference exists between the energy region
$\sqrt{q^2}\simeq 1~ TeV$ (possibly within the final aimed reach of LC)
and that of $\sqrt{q^2}\simeq 3-4~ TeV$ (possibly within the CLIC
range). In the first case, the same previous conclusions about the
reliability of a one-loop expansions are still, in our working
assumptions, essentially valid (in fact, the relative difference
between the one-loop and the resummed effects is always below the
assumed visible one percent level. This reliability is totally lost
when one reaches the $\simeq3~TeV$ regime. In this case, the relative
difference between the two effects is well beyond the one percent
level, particularly for large $\tan\beta$. In the extreme case of Higgs
boson production at $\tan\beta=40$, the relative difference between the
two effects is of approximately five percent at $3~TeV$. Similar
features are valid for squark production as well.
Thus, for third family squarks and charged Higgs boson production in the
CLIC energy regime, with a SUSY mass of few hundred TeV, stopping the
theoretical calculation at one loop level would be in our opinion a
theoretical catastrophe. It should also be stressed that the resummed
effect remains large and sometimes spectacular. In particular, in the
extreme case of charged Higgs bosons production at $\tan\beta=40$, it
reaches the 35 percent value at $\sqrt{q^2}=3~TeV$. For squarks, the
effect is reduced but is still generally large (from $\simeq5$ to
$\simeq15$ depending on the cases). We insist again on the fact that,
as a consequence of the color factor in the quark loop in Fig.2c, the
process of charged Higgs production exhibits the most sizeable SUSY
Yukawa Sudakov effect, as we anticipated in Sect.II.\par
In the case of forward-backward asymmetries, depicted in 
Figs.\ref{afbstl},\ref{sigmaafbhiggs}b ,
the situation is very similar to that of the sleptons and first
families of squarks, as one can see. The only difference appears in the
case of Higgs boson production. Here the difference between one-loop
and resummed effects becomes again visible in the CLIC (but not in the
LC) regime. This fact is, in our opinion, accidental since we could
not find deeper physical motivations for it.\par
The \underline{second} relevant feature of 
Figs.\ref{tanbetastl},\ref{sigmaafbhiggs}a is the fact that
a strong dependence on $\tan\beta$ appears in the logarithmic effect.
This can be seen by comparison of the two curves that correspond to
$\tan\beta=10$ and $\tan\beta=40$. As one notices, the relative
\underline{difference} between the two effects is always "large"
(depending on the case again with an enhancement in the Higgs boson
case), with the only exception of right stop production (this can be
qualitatively understood looking at Eq.(\ref{cfsur})). 
This remarkable feature
remains essentially true in the overall $1-4~TeV$ energy range,
although it increases with energy. The previous observation suggests
that, from a special analysis of cross sections, one might be able to
derive interesting informations on $\tan\beta$. In fact, this
possibility was already considered in a previous paper devoted to top
quark production \cite{topq}, where it was proposed to exploit
measurements of the \underline{slope} of the cross section to fix the
$\tan\beta$ value. In the next Sect.V we shall generalize the previous
proposal to the cases of stop, sbottom, charged Higgs boson production,
and discuss in some detail the possible consequences for a relative
precise determination of $\tan\beta$.\\

A final comment has to be added concerning the mixing effects which
affect the third family of squarks. The mass eigenstates will no more be
$\tilde{f}_{L,R}$ but the combinations
$\tilde{f_1} = \cos\theta_f\tilde{f_L}+ \sin\theta_f\tilde{f_L}$
and $\tilde{f_2} = -\sin\theta_f\tilde{f_L}+ \cos\theta_f\tilde{f_L}$.
Experimental results will be obtained
for production of $\tilde{f_1}\tilde{\bar{f_1}}$,
$\tilde{f_2}\tilde{\bar{f_2}}$, $\tilde{f_1}\tilde{\bar{f_2}}$,
$\tilde{f_2}\tilde{\bar{f_1}}$. 
However in the asymptotic regime regime,
the amplitudes for mass eigenstates
can be expressed in terms of amplitudes for chiral $L,R$ states
in a simple way:

\bqa
&&A_{11}=A_{LL}\cos^2\theta_f+A_{RR}\sin^2\theta_f\nonumber\\
&&A_{22}=A_{LL}\sin^2\theta_f+A_{RR}\cos^2\theta_f\nonumber\\
&&A_{12}=A_{21}=(A_{RR}-A_{LL})\sin\theta_f\cos\theta_f
\eqa
\noindent
so that it should be straightforward to express the experimental results
in terms of the observables concerning the chiral states that we 
considered in this paper. One has just to invert the above equations,
and one obtains:

\bq
tan2\theta_f={2A_{12}\over A_{22}-A_{11}}
\eq
\bq
A_{LL}={A_{11}\cos^2\theta_f-A_{22}\sin^2\theta_f
\over\cos^2\theta_f-\sin^2\theta_f} 
\eq
\bq
A_{RR}={A_{22}\cos^2\theta_f-A_{11}\sin^2\theta_f
\over\cos^2\theta_f-\sin^2\theta_f} 
\eq
\noindent
for each flavor $\tilde{t}$ or $\tilde{b}$, separately.
The experimental results can then be given for chiral
states and compared with the theoretical predictions
made throughout this paper.

\section{Determination of $\tan\beta$}
\label{sectanbeta}

In the framework of the MSSM, we showed in a previous paper~\cite{14} that 
the cross section for the process $e^+e^-\to q\overline{q}$, 
with $q$ being a third generation quark (top and bottom), 
contains angular independent Sudakov logarithms of Yukawa origin.
These are terms 
depending on $\tan\beta$ and on the SUSY mass $M_{SUSY}$, 
which are the only
SUSY parameters surviving in the high energy limit of this process.

To understand the reason for this peculiar feature we recall that 
the free 
parameters of the MSSM can be broadly divided into three classes: 
(i) the ones belonging to the SM sector, (ii)
$\tan\beta$ that is related to the two doublet structure of the 
Higgs sector and (iii) the (many) SUSY soft breaking mass terms.

As we discussed in~\cite{14}, an analysis of the slope of the effects in the
observables as the energy is increased allows to extract the 
value of $\tan\beta$ without any 
specific knowledge of the other parameters. This very welcome feature 
is present also in the processes considered in this paper;
to be more precise, it remains rigorously true when working in the 
one loop approximation, and it is valid to subleading order accuracy
if one uses the complete resummed expressions.
The remaining
part of this Section will be devoted to a numerical analysis 
with the specific aim of determining $\tan\beta$.

Let us denote by $\sigma_n$, $n=1-5$ the various cross sections for production 
of $\widetilde{t}_{L, R}$, $\widetilde{b}_{L, R}$ or charged Higgs 
bosons $H^\pm$. We define the relative SUSY effect on the cross section
$\sigma_n$ as the ratio
\bq
\epsilon_n(q^2) = \frac{{\cal O}_n(q^2)-
{\cal O}^{SM}_n(q^2)}{{\cal O}^{SM}_n(q^2)} .
\eq
This definition is useful as far as we can regard the SM contributions as 
perfectly known terms. As we have already discussed,
at energies around 1 TeV, this statement is certainly true because
a one loop calculation is perfectly reliable and reproduces 
the full effect, with resummation giving a negligible additional shift
in the observables.

At one loop, in the asymptotic regime, 
we can parametrize $\epsilon_n$ as 
\bq
\label{expansion}
\epsilon_n(q^2) = F_n(\tan\beta) \ln\frac{q^2}{M_S^2} + G_n + \cdots .
\eq
Here, $F_n$ is a function of $\tan\beta$ only. The explicit form of 
its $\tan\beta$ dependent terms can be 
obtained from the Yukawa terms and we write them here for clarity and
convenience of the reader
\bq
F_{\widetilde{t}_L} = -\frac\alpha\pi\ \frac{1}{4M_W^2 s_W^2}\left(\frac{m_t^2}{\tan^2\beta}
+m_b^2\tan^2\beta\right) ;
\eq
\bq
F_{\widetilde{t}_R} = -\frac\alpha\pi\ \frac{1}{2M_W^2 s_W^2}\ \frac{m_t^2}{\tan^2\beta} ;
\eq
\bq
F_{\widetilde{b}_L} = F_{\widetilde{t}_L} ;
\eq
\bq
F_{\widetilde{b}_R} = -\frac\alpha\pi\ \frac{1}{2M_W^2 s_W^2}\ m_b^2 \tan^2\beta ;
\eq
\bq
F_{H^\pm} = -\frac\alpha\pi\ \frac{3}{4M_W^2 s_W^2}\left(\frac{m_t^2}{\tan^2\beta}
+m_b^2\tan^2\beta\right) .
\eq
The constant $G_n$ in Eq.~(\ref{expansion}) is a sub-subleading correction.
It does not increase with $q^2$, 
but depend on all mass ratios of virtual particles.
The omitted terms in Eq.~(\ref{expansion}) vanish in the high energy limit.

To eliminate $G_n$ we proceed as in~\cite{14}. We assume 
that a set of $N$ independent measurements is available 
at c.m. energies $\sqrt{q_1^2}, \sqrt{q_2^2}, \dots,\sqrt{q_N^2}$
and take the difference of each measurement 
with respect to the one at lowest energy. The resulting quantities
\bq
\delta_{n, i} = \epsilon_n(q_i^2)-\epsilon_n(q_1^2) ,
\eq
do not contain $G_n$ and take the simple form
\bq
\delta_{n, i} = F_n(\tan\beta^*) \ \ln\frac{q_i^2}{q_1^2} ,
\eq
where $\tan\beta^*$ is the {\em true} unknown value that describes the
experimental measurements.

We now describe how precisely $\tan\beta$ can be extracted.
We denote by $\sigma_n(q^2)$ the experimental error on $\epsilon_n(q^2)$.
For each set of explicit measurements $\{\delta_n(q_i^2)\}$, 
the best estimate for $\tan\beta$ is the value that minimizes 
the $\chi^2$ sum
\bq
\chi^2(\tan\beta) = \sum_{i=1}^N\sum_{n=1}^{N_{\cal O}}
\frac{[F_n(\tan\beta) 
\ln\frac{q^2_{i+1}}{q^2_1} - \delta_{n, i}]^2}{4\sigma_{n, i}^2} ,
\eq
where $\delta_{n, i} \equiv \delta_n(q^2_i)$ and $\sigma_{n, i} \equiv \sigma_n(q_i^2)$.

\begin{figure}[htb]
\vspace*{2cm}
\[
\epsfig{file=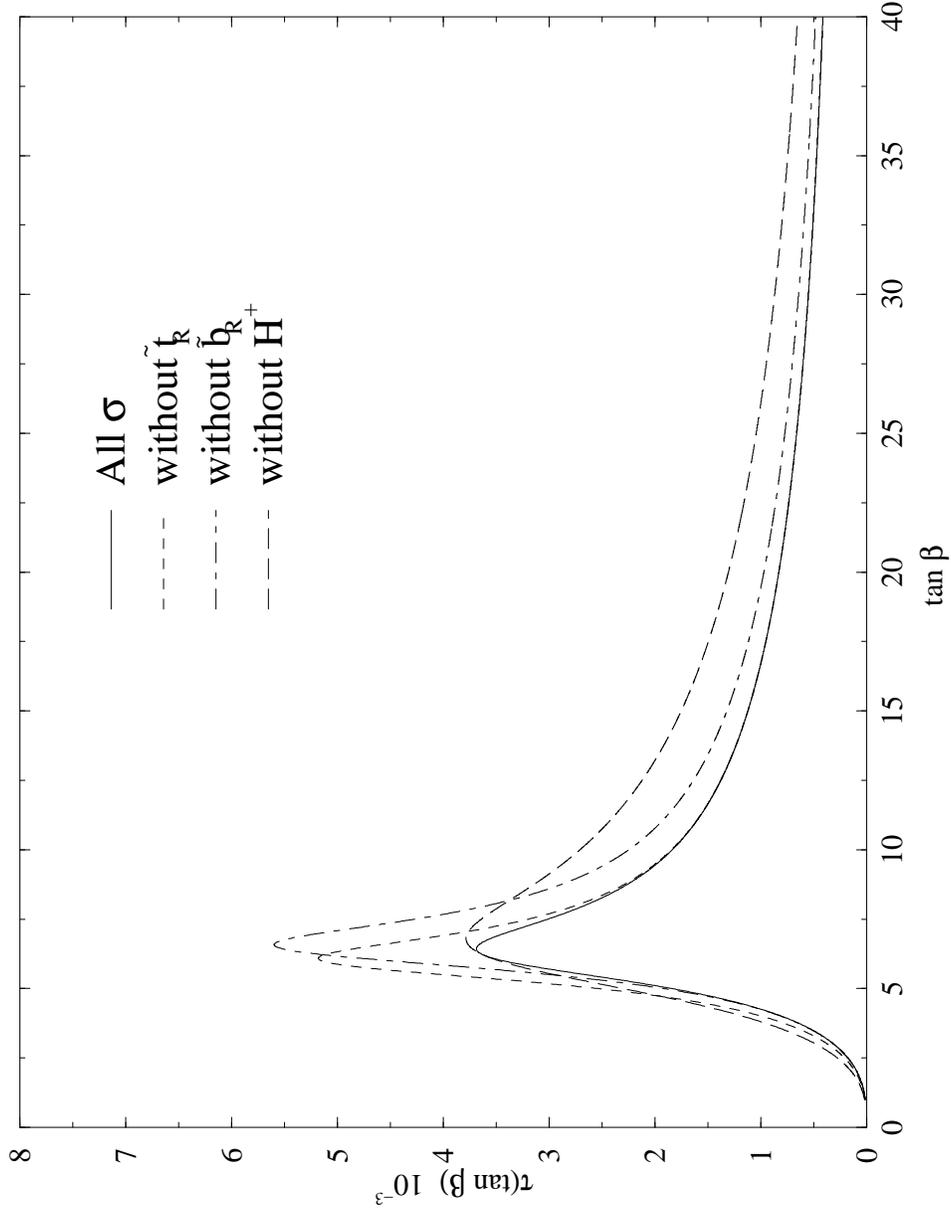,height=16cm}
\]
\vspace*{1cm}
\caption[1]{This Figure shows the behavior of the $\tau$ function 
defined in Eq.~(\ref{taudef}). It increases when the slope of the 
SUSY effects does not depend much on $\tan\beta$. The four lines
correspond to the four possible choices discussed in 
Sec.~(\ref{sectanbeta}), that is: (i) all the five cross sections for 
production of third generation squarks and charged Higgs bosons,  
(ii) without production of $\widetilde{t}_R$, (iii)
without production of $\widetilde{b}_R$, (iv)
without production of charged Higgses $H^\pm$.
}
\label{tau}
\end{figure}

As usual, 
the experimentally measured quantity $\delta_{n, i}$ is assumed to be a 
normal Gaussian random variable distributed around the value
\bq
\label{diff}
F_n(\tan\beta^*) \ln\frac{q^2_{i+1}}{q^2_1} ,
\eq
with standard deviation $2\sigma_{n,i}$.  
After linearization around $\tan\beta=\tan\beta^*$, 
minimization of $\chi^2$ provides the best estimate of $\tan\beta$.
This is an unbiased Gaussian estimation with  mean 
$\tan\beta^*$ and standard deviation fixed by the condition 
$\Delta\chi^2=1$, i.e. 
\bq
\delta\tan\beta = 2\left(\sum_{n, i} \left(\frac{F_n'(\tan\beta^*) 
\ln\frac{q_{i+1}^2}{q_1^2}}
{\sigma_{n,i}}\right)^2\right)^{-1/2} .
\eq
Under the simplifying assumption $\sigma_{n, i} \equiv \sigma$, this
formula reduces to 
\bq
\delta\tan\beta = 2\sigma\left(\sum_n F_n'(\tan\beta^*)^2\right)^{-1/2}
\left(\sum_i \ln^2\frac{q_{i+1}^2}{q_1^2} \right)^{-1/2} .
\eq
The function 
\bq
\label{taudef}
\tau(\tan\beta) = \left(\sum_n F_n'(\tan\beta)^2\right)^{-1/2} ,
\eq
measures the dependence of the slope of SUSY effects on $\tan\beta$. It is 
shown in Fig.\ref{tau} 
for four possible choices: (i) all the five cross sections, (ii)
without production of $\widetilde{t}_R$, (iii)
without production of $\widetilde{b}_R$, (iv)
without production of charged Higgses $H^\pm$.

In the best case (i), it is strongly peaked around $\tan\beta = 6$
and the combination of the various observables, 
especially the cross sections for production of right sfermions 
(the ones with larger $\cot^2\beta$ coefficient) is crucial to keep 
the function $\tau(\tan\beta)$ as small as possible.

\begin{figure}[htb]
\vspace*{2cm}
\[
\epsfig{file=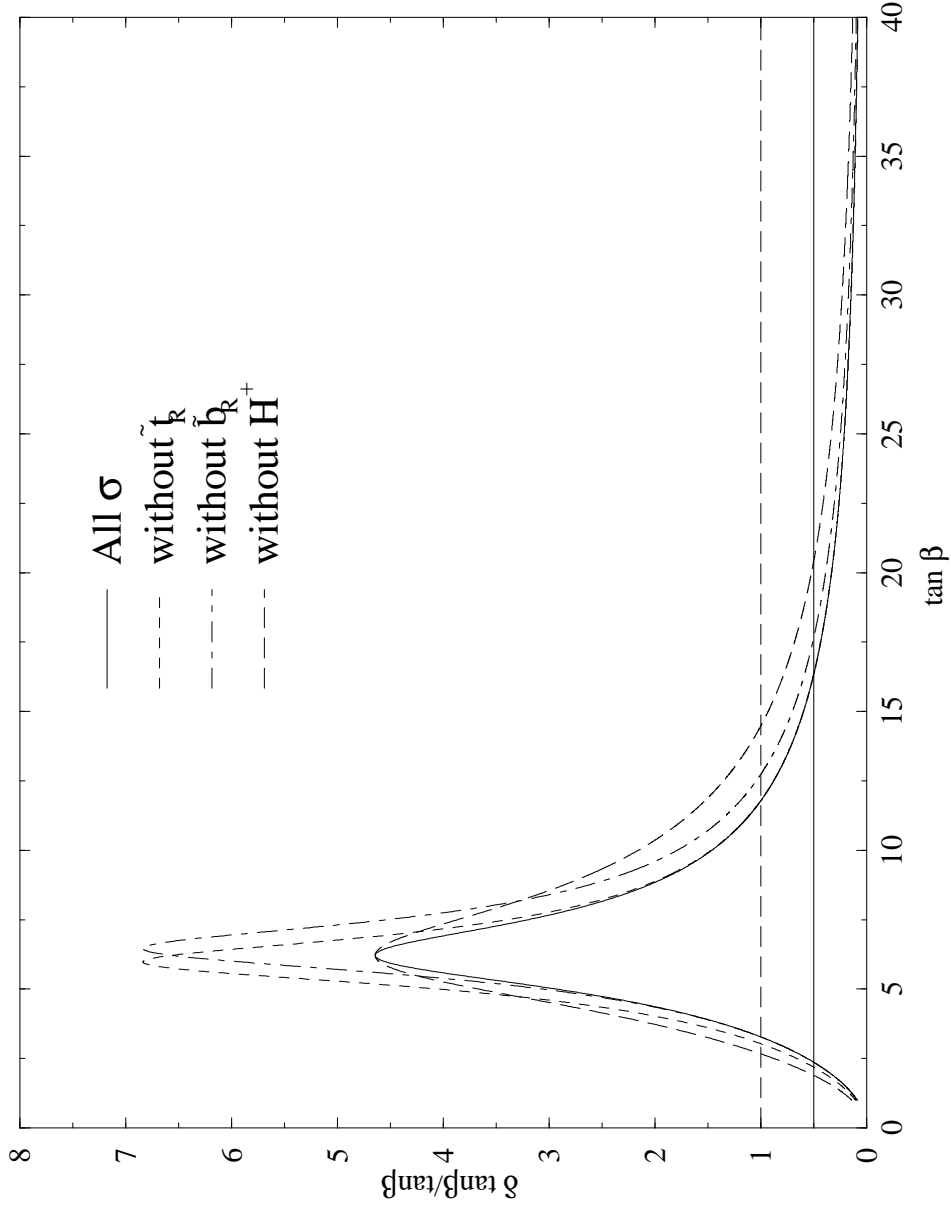,height=16cm}
\]
\vspace*{1cm}
\caption[1]{
We plot in this Figure the relative error $\delta\tan\beta/\tan\beta$ 
that can be derived in the determination of $\tan\beta$
assuming a relative accuracy of 1\% on all the cross sections and 
the availability of 10 measurements at equally spaced energies 
between 800 GeV and 1.5 TeV.
for all the five observables.
Again, we consider the optimal scenario when all the observables can
be exploited as well as what happens when a subset of them is removed.
In the best case, values of $\tan\beta > 16$ can be determined
with a 50\% relative accuracy.
}
\label{error}
\end{figure}

To understand the consequences of the shape of $\tau$, we plot in 
Fig.\ref{error} the relative error $\delta\tan\beta/\tan\beta$ 
computed under the optimistic assumption of a relative accuracy 
equal to 1\% for all the five observables.
The three curves correspond to the assumption that 
independent 
measurements for each observable are available at $N=10$  
equally spaced c.m. energies around 1 TeV,  between 0.8 TeV and 1.5 TeV.
We remark that different curves associated to pairs $(N, \sigma)$ 
depend only on the combination $\sigma/\sqrt{N}$.
In the Figure, we also show horizontal dashed lines corresponding to relative errors
equal to 1 and 0.5. As one can see in the Figure, values in the range
\bq
\tan\beta < 3, \qquad \tan\beta > 16
\eq
can be detected with $N=10$ c.m. energy values
with a relative error smaller than 50\%, that can be  considered
qualitatively as a ``decent'' accuracy. If a higher experimental
precision (e.g. a few permille in the cross sections) were achievable, 
the same result could be obtained with a smaller number ($N\simeq 3$) of 
independent energy measurements. The first region ($\tan\beta<3$) appears
to be unfavoured by the present LEP combined data analysis, but the second
one ($\tan\beta>14$) is instead very interesting for the LHC physics 
programme.

The extension of the previous analysis to 
the case of a Linear Collider working at an energy around 
3 TeV (CLIC), is not straightforward.
In fact, 
the one loop expressions are no more reliable and a resummation of higher order
terms must be performed. 
Up to date, the best theoretical accuracy that can be reached in the 
MSSM is the subleading one discussed in this paper. 
The Yukawa terms are then given at all orders by combining the 
one loop Yukawa contributions with the resummed double logarithms
of gauge origin. To this level of accuracy, the relative 
SUSY Yukawa effect is thus unchanged and the analysis can be repeated
with the same formulae.
However, one must keep in mind that the subleading approximation can 
be enough to determine the gross size of the virtual 
effects, but its validity must be checked in the analysis of 
finer details like the dependence on $\tan\beta$ of the omitted constant terms.
Notwithstanding these unavoidable remarks, we can analyze what happens at
3 TeV in the subleading approximation. 

As we said above, the same expressions as in the 1 TeV analysis can be 
applied. The need to eliminate any sub-subleading constant
from the SUSY effects leads again to Eq.~(\ref{diff}).
The effects thus depend on the logarithms of the ratios between the 
various measurement energies and the lowest energy $q_1$. 
Therefore the same results on the error $\delta\tan\beta/\tan\beta$ 
that we derived around 1 TeV can be extended to the rescaled 
(wider) energy range 
\bq
[0.8,1.5]\  {\rm TeV}\to [2.7, 4.5]\ {\rm TeV}
\eq
with no additional remarks or changes.
 
\begin{figure}[htb]
\vspace*{2cm}
\[
\epsfig{file=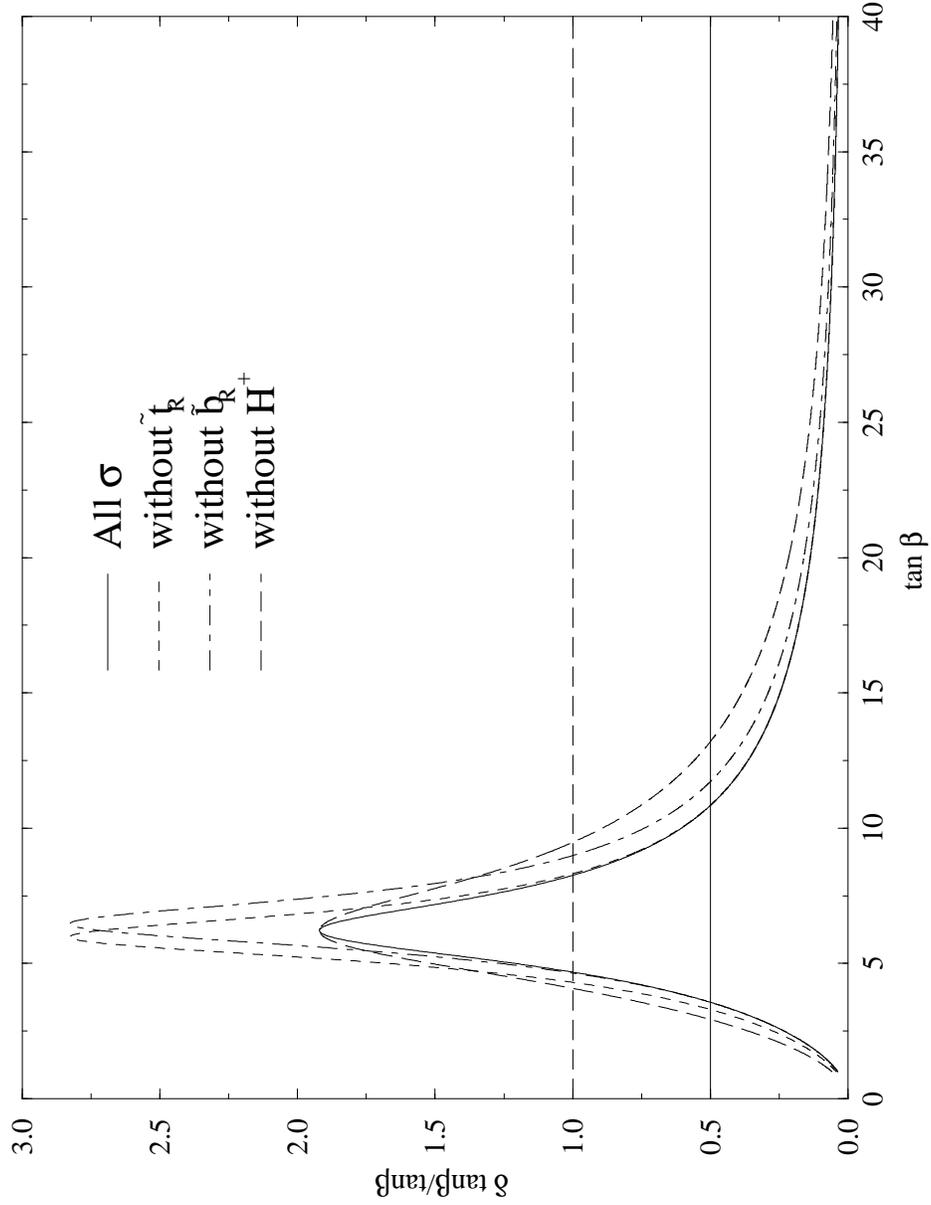,height=16cm}
\]
\vspace*{1cm}
\caption[1]{
Relative error $\delta\tan\beta/\tan\beta$ as in the previous 
Figure, but assuming the availability of 
10 measurements at equally spaced energies 
between 800 GeV and 3.3 TeV (and again a relative 1\% error on the 
measurements).
In the best case, values of $\tan\beta > 11$ can be determined
if the relative accuracy is around 50\%, $\tan\beta > 14$ if it is 25\%.
}
\label{error3}
\end{figure}

It might be interesting to discuss what would be the change in 
such an analysis if data could be accumulated starting at 800 GeV and
increasing the energy up to an upper value around 3 TeV. 
In this case, shown in Fig.\ref{error3}, the results are largely
improved with respect to those that could be
with measurements at  1 or 3 TeV only. 
To be more precise, with $10$ measurements extending 
from 800 GeV to 3.3 TeV, the region $\tan\beta > 11$ can be 
accessible with 50\% accuracy. This bound reduces to $\tan\beta > 14$ if
the required accuracy is 25 \%. 
We stress that this is a quite interesting region
\footnote{For completeness, 
we remark that here too, 
as in the previous analysis, there is a region at low $\tan\beta$ 
that can be analyzed.} 
as discussed in~\cite{15},
where it is shown that a measurement of $\tan\beta$ 
is practically impossible from chargino
or neutralino production when $\tan\beta>10$ 
since the effects depend on 
$\cos2\beta$ that becomes flat for $\beta\to \pi/2$. 
It could be achieved 
in the associated productions $e^+e^-\to h\tilde\tau\tilde\tau$ or 
$e^+e^-\to A\bar{b}b$ (with $h$ and $A$ being the CP even and odd Higgs 
bosons), but only for very large $\tan\beta$ values ($\sim 50 $).

\section{Conclusions}

In this final Section we shall draw a number of 
relevant conclusions. Before doing that, we feel though, that some
preliminary considerations are opportune.\par
In this paper we have derived the complete Sudakov one-loop expansion
and the related all orders subleading resummation for SUSY scalar production.
Since both expressions have been computed for the first time, there are
no different papers in the literature with which to compare our results.
It is, however, a highly non-trivial check that all SM DL and SL terms (in particular
the angular dependent contributions) are in agreement with the results
obtained in the symmetric basis in Sect.III. 
For the SL terms of genuine supersymmetric origin obtained from  
the physical one-loop calculation that we
have performed, and that has led us to a large number of equations and
formulae, no such internal cross check is available.\par
In spite of the lack of possible comparisons, however, there are a few other internal
consistency checks of our results that serve to
support their validity. Here we list a limited number of them in the
following order:\\

a) The box diagrams with $2W$ exchange and the vertex diagrams with
initial electron and final scalar $2W$ triangle get combined separately
in a tricky way. One half of the box must be summed with the initial
$2W$ vertex, the other one with the final $2W$ vertex. After these
summations, one gets the correct universal factors $L_3=(3\ln -\ln^2)$
for the initial state and $L_4=(4\ln -\ln^2)$ for the final state,
although none of the separate ($WW$ box, $WW$ initial vertex, $WW$
final vertex) contributions produces alone the $L_3,~L_4$ terms.\\

b) The overall MSSM "gauge" effect for the final scalar pair, obtained
by summing the diagrams with SM gauge bosons (i.e. photon,$Z$,$W$) to
those with SUSY gauginos, is proportional to a \underline{new}
combination $L_2=(2\ln -\ln^2)$. This is the \underline{same} MSSM
"gauge" combination, with the \underline{same} coefficient that affects
the \underline{respective} fermionic superpartners either in the
initial or in the final state of $e^+e^-$ annihilation. One can easily
check this statement for final sleptons looking at
eqs.(\ref{cingIY}),(\ref{cinsIY}),
(\ref{cfing}-\ref{cfslr}) of this paper. For final quarks, 
one should compare
the expressions of this paper with the analogous ones given for quark
production in the MSSM in a previous paper \cite{13} reducing these
expressions to the opportune form (which shall not be explicitly shown
here). This equality is expected, a priori, since the contributions
are computed for vanishing particles-sparticles masses, where
supersymmetry is supposed to be exact. As a consequence, the pure
overall gauge effects should be the same within the same
supermultiplet.\\

c) In the MSSM, the final \underline{Yukawa} effects for stop and
sbottom production at lepton colliders are the same as those for top
and bottom production. This can be also verified by comparing the
expressions of this paper with the corresponding ones of previous
references \cite{13,14}. This equality is highly non trivial in a
technical sense, since in the two cases quite different diagrams
contribute (in particular, virtual Higgs bosons vertices do not
contribute for squark production, while the corresponding SM scalars
\underline{do} contribute for top, bottom final pairs). We interpret it
as a consequence of the MSSM origin of Yukawa couplings, carried by the
Higgsino, which is the SUSY superpartner of the Higgs boson and lies
therefore in the same supermultiplet.\par

Having shown the three non trivial consistency checks of our one-loop
calculations, we now proceed assuming that our results are correct and
draw some final conclusions. Briefly, first of all it seems to us that
the process of SUSY scalar production in its Sudakov regime, assuming
that the latter coincides with the TeV energy range for reasonably
light SUSY masses, can be divided into two distinct categories. The
first one includes slepton and first two family squark production. 
The second one includes the considered
third family squark and charged Higgs boson production. For all
cases, there is an energy dependence of the one-loop approximation,
which begins to be unreliable when one crosses the $\simeq$ 1,2
 TeV
region (where it should still be satisfactory). At CLIC energies, the
need of a proper resummation becomes imperative and all our one-loop
results must be replaced by our corresponding resummed expressions. The
important point is that in this procedure, the (generally) large
unreliable one-loop effects are replaced by reliable resummed effects,
that are sensibly reduced but still remain large and visible given the
expected level of experimental accuracy. Independently of this fact, a
strong $\tan\beta$ dependence in the third family and
charged Higgs boson production case
(particularly effective in the charged
Higgs boson case) should allow to achieve, via suitable measurements of
the slope of the cross sections in the total energy range $1-4~TeV$ a
satisfactory determination of the SUSY parameter $\tan\beta$. This should
be combined with other analogous independent measurements to be 
performed in other processes (like e.g. neutral Higgs boson production)
to obtain a more precise determination (an analysis of this possibility
is already in progress\cite{neut}).\par
Finally, we believe to have shown in this example that there exists a
realistic energy range, to be hopefully covered in a not too far
future, where virtual corrections of the MSSM can become
\underline{large}. Given the previous experience at lower energies in
the hundred GeV where SUSY virtual effects were systematically small,
we believe that this fact would provide a clean and fundamental test of
the validity of the MSSM at future lepton colliders, in case
supersymmetric particles were detected somewhere in a (hopefully) near
future.


\begin{table}
\caption{Numerical values of the $k$ coefficients. They are defined
in Eq.~(\ref{kdef}) and represent the relative weights of the various
logarithmic one loop effects in the cross sections.
The second part of the table shows their numerical values in the case
of the $A_{FB}$ asymmetry. A dash denotes a contribution that is 
rigorously null and not simply numerically negligible.}
\vskip 0.5cm
\begin{tabular}{l|rrrrrrr}
100 $\Delta\sigma/\sigma$
& $k^{RG}_{SM}$ & $k^{RG}_{SUSY}$ & $k^{in}$ & 
$k^{fin}_{gauge}$ &  
$k^{Yuk}_t$ & $k^{Yuk}_b$ & $k^{Box}$ \\
\\
\tableline
\\
$\widetilde{t}_L$ & -1.3 & 1.9 & 0.41  & 0.38 & -0.25 & -0.25 & 1.5 \\
$\widetilde{t}_R$ & 1.0  & 0.63& 0.20  & 0.067& -0.51 & --    & --  \\
$\widetilde{b}_L$ & -1.7 & 2.2 & 0.40  & 0.38 & -0.25 & -0.25 & 1.8 \\
$\widetilde{b}_R$ & 1.0  & 0.63& 0.20  & 0.017& --    & -0.51 & --  \\
$H^\pm$           & -0.63& 1.6 & 0.37  & 0.41 & -0.76 & -0.76 & 1.1 \\
\\
\tableline
\\
$\Delta A_{LR}$ & $k^{RG}_{SM}$ & $k^{RG}_{SUSY}$ & $k^{in}$ & 
$k^{fin}_{gauge}$ & 
$k^{Yuk}_t$ & $k^{Yuk}_b$ & $k^{Box}$ \\
\\
\tableline
\\
$\widetilde{t}_L$ & -0.15 & 0.083 & 0.016 & -- & -- & -- & 0.095   \\
$\widetilde{t}_R$ & --    & --    & 0.084 & -- & -- & -- & -- \\
$\widetilde{b}_L$ & -0.26 & 0.15  & 0.024 & -- & -- & -- & 0.17 \\
$\widetilde{b}_R$ & --    & --    & 0.084 & -- & -- & -- & -- \\
$H^\pm$           & -0.59 & 0.33  & 0.076 & -- & -- & -- & 0.37
\end{tabular}
\label{tab1}
\end{table}

\begin{table}
\caption{Absolute shift in the asymmetry $A_{LR}$. We consider here
two energies $1$ and $3$ TeV. For each energy and final sparticle
we show values. ``1L'' is the one loop contribution, ``Res'' is the 
contribution resummed at subleading accuracy. The label ``no $\theta$''
means that the {\em genuine} non universal angular contributions from boxes
have been suppressed. About the sparticle labeling, we remind that 
$\widetilde{l}$ stands for a generic slepton $\widetilde{\mu}$ or
$\widetilde{\tau}$ and that $\widetilde{u}$ and $\widetilde{d}$ stand  
for squarks with $T^3=1/2$ and $-1/2$ in the first two generation.}
\vskip 0.5cm
\begin{tabular}{l|rrrrrrrr}
100 $\Delta A_{LR}$ & 1 TeV & 1 TeV & 1 TeV           & 1 TeV  & 
                      3 TeV & 3 TeV & 3 TeV           & 3 TeV  \\
                    & 1L    & Res  & 1L, no $\theta$  & Res, no $\theta$
                    & 1L    & Res  & 1L, no $\theta$  & Res, no $\theta$ \\
\\
\tableline
\\
$\widetilde{l}_L$ &-0.27  & -0.17   & -2.4  & -2.3  & -1.9  & -1.6  & -5.7 & -4.8 \\
$\widetilde{l}_R$ &-1     & -0.99   & -1    & -0.97 & -3    & -2.6  & -3   & -2.5 \\
$\widetilde{\nu}$ & 0.53  &  0.89   & -6    & -5.9  & -1.5  & -1.3  & -13  & -11  \\
$\widetilde{u}_L$ &-0.031 & -0.0069 & -0.57 & -0.56 & -0.36 & -0.31 & -1.4 & -1.2 \\
$\widetilde{u}_R$ &-1     & -0.98   & -1    & -0.97 & -2.9  & -2.6  & -2.9 & -2.5 \\
$\widetilde{d}_L$ & 0.011 &  0.055  & -0.95 & -0.93 & -0.43 & -0.37 & -2.2 & -1.9 \\
$\widetilde{d}_R$ &-1     & -0.98   & -1    & -0.97 & -2.8  & -2.6  & -2.8 & -2.6 \\
$\widetilde{t}_L$ &-0.032 & -0.017  & -0.59 & -0.57 & -0.39 & -0.34 & -1.5 & -1.2 \\
$\widetilde{t}_R$ &-1.1   & -0.98   & -1.1  & -0.97 & -3.2  & -2.5  & -3.2 & -2.5 \\
$\widetilde{b}_L$ & 0.011 &  0.037  & -0.99 & -0.95 & -0.46 & -0.43 & -2.4 & -2   \\
$\widetilde{b}_R$ &-1     & -0.97   & -1    & -0.97 & -2.9  & -2.6  & -2.9 & -2.6 \\
$H^\pm$           &-0.3   & -0.3    & -2.6  & -2.4  & -2.4  & -2    & -7.3 & -5.2
\end{tabular}
\label{tab2}
\end{table}

\newpage

\newpage

\end{document}